\documentclass[twocolumn,twocolappendix]{aastex631}
\usepackage{amsmath,amssymb,bm}
\usepackage{graphicx}

\newcommand{\vect}[1]{\mbox{\boldmath{${#1}$}}}
\newcommand{\sub}[1]{_{\mbox{\scriptsize {#1}}}}
\newcommand{\superscript}[1]{^{\mbox{\scriptsize {#1}}}}

\received{March 13, 2024}
\revised{April 13, 2021}
\accepted{\today}

\shorttitle{Three-dimensional Interaction between a Planet and an 
Isothermal Gaseous Disk}
\shortauthors{Tanaka and Okada}

\begin{document}

\title{
Three-dimensional Interaction between a Planet and an Isothermal 
Gaseous Disk.\\
III. Locally Isothermal Cases
}

\correspondingauthor{Hidekazu Tanaka}
\email{hidekazu@astr.tohoku.ac.jp}

\author[0000-0001-9659-658X]{Hidekazu Tanaka}
\affiliation{Astronomical Institute, Graduate School of Science, Tohoku University, 
6-3, Aramaki, Aoba-ku, Sendai 980-8578, Japan}

\author{Kohei Okada}
\affiliation{Astronomical Institute, Graduate School of Science, Tohoku University, 
6-3, Aramaki, Aoba-ku, Sendai 980-8578, Japan}
\affiliation{IBM Japan Ltd., 19-21, Nihonbashi-Hakozakicho, Chuo-Ku, Tokyo, 13 103-8510, Japan}



\begin{abstract}
We performed linear calculations to determine the Type I planetary migration 
rate for three-dimensional locally isothermal disks with radial temperature 
gradients. For 3D disks with radial temperature gradients, the linear wave equation 
has a {divergent term of the third pole, which makes corotation a non-removal singularity.}
We suppressed the divergence with the Landau prescription to obtain the wave solutions. 
Despite the {singularity at corotation}, we derived a definite torque on the 
planet because the divergent term amplifies the waves only in the neighborhood of 
corotation and has little effect on the planetary torque. Consequently, we derived 
the formulas for the total, Lindblad, and corotation torques for locally isothermal 
disks. The resulting torque term due to the disk temperature gradient agrees well with 
the results of previous 3D hydrodynamical simulations for locally isothermal disks. 
Our linear calculation also provides the 3D horseshoe torque,  which is close to the 
results of previous 3D hydrodynamical simulations.
\end{abstract}

\keywords{Planet formation(1241) --- Protoplanetary disks(1300)
--- Planetary Systems: planet–disk interactions
}


\section{Introduction}\label{sec:intro}
Type I planetary migration is a crucial process in the planet formation 
theory. It is caused by the gravitational interaction between a gapless gas
disk and an embedded low-mass planet. Initially, the Type I migration had 
been studied with two-dimensional linear calculations 
\citep[e.g.,][]{1979ApJ...233..857G,1980ApJ...241..425G,1979MNRAS.186..799L,
1986Icar...67..164W,1993ApJ...419..166A,1993Icar..102..150K}, and 
three-dimensional calculations were later carried out by \citet{1998PASJ...50..141T} and 
\citet[hereafter referred to as Paper I]{2002ApJ...565.1257T}.
The Type I migration torque arises from Lindblad and corotation resonances.
In locally isothermal disks with moderate density and temperature gradients, 
the torque due to Lindblad resonances generally overcomes that due to
corotation resonances and causes inward planetary migration.
In Paper I, a torque formula was provided 
for three-dimensional isothermal disks with a surface density of 
$\sigma \propto r^{-\alpha}$ as
\begin{equation}
\Gamma = -(1.36+0.54\alpha) q^2 \left( \frac{h_p}{r_p} \right)^{-2}
\sigma_p r_p^4 \Omega_p^2, 
\end{equation}
where $q$ is the mass ratio of the planet to the host star, 
$r_p$ and $\Omega_p$ are the orbital radius and angular velocity 
of the planet, and $h_p$ and $\sigma_p$ are the disk scale height 
and surface density at $r=r_p$.
This torque formula gives a short inward migration time of 
$\sim10^5$yrs for a Jovian solid core with 10M$_\oplus$ at 
5AU in a typical protoplanetary disk with a surface density of 
100g\,cm$^{-2}$ and a temperature of 100K. Rapid Type I migration 
is a major obstacle to the formation of solid cores of 
giant planets due to the much longer core accretion time
of $\sim 10^7$yrs \citep[e.g., ][]{2002ApJ...581..666K}.
Many attempts have been made to accelerate the core accretion
with a significant supply of pebbles or tiny planetesimals 
\citep[e.g.,][]{2010A&A...520A..43O,2012A&A...544A..32L,
2021ApJ...922...16K,2023ApJ...954..158K}.

Several three-dimensional hydro-dynamical simulations have shown
good agreements with the migration rates given by this torque formula
\citep[e.g.,][]{2003ApJ...586..540D,2003MNRAS.341..213B,
2010ApJ...724..730D}. These three-dimensional hydrodynamical simulations 
have also been done for locally isothermal disks with non-zero 
temperature gradients. \citet{2010ApJ...724..730D} have systematically
investigated the dependence of the torque on the disk temperature 
gradient as well as the dependence on the surface density gradient. 
Their results have been confirmed by \citet{2017MNRAS.471.4917J}. 
However, three-dimensional linear calculations of the Type I migration 
torque have not yet been performed for radially non-isothermal disks.

In the last two decades, research on corotation torques has made tremendous
progress through hydrodynamical simulations and linear calculations
\citep{2001ApJ...558..453M,2006A&A...459L..17P,2008ApJ...672.1054B,
2008A&A...485..877P,2009MNRAS.394.2283P,2009MNRAS.394.2297P,
2009ApJ...703..857M,2010ApJ...723.1393M,2010MNRAS.401.1950P,
2011MNRAS.410..293P,2011A&A...536A..77B,2014MNRAS.444.2031P,
2014MNRAS.440..683L,2016ApJ...817...19M,2017MNRAS.471.4917J,
2017AJ....153..124F,2023ASPC..534..685P}.
These studies have shown that in adiabatic or
radiative disks, the nonlinear corotation torque (or the so-called 
horseshoe torque) is strongly enhanced and can drive outward migration 
for a certain range of planetary masses. Based on two-dimensional
hydrodynamical simulations, \citet{2010MNRAS.401.1950P,2011MNRAS.410..293P}
developed a new torque formula that includes the nonlinear horseshoe 
torque. \citet{2010ApJ...723.1393M}
also independently constructed a similar two-dimensional torque formula.
These two-dimensional torque formulas depend on the softening parameter.
\citet{2017MNRAS.471.4917J}
systematically performed 3D hydrodynamical simulations and derived the improved 
3D torque formula.

While linear calculations are useful in verifying hydrodynamical simulations, 
the three-dimensional linear calculation of the Type I migration has only been 
conducted for completely isothermal disks. We demonstrated in Paper I that the 
linear wave equation for three-dimensional density waves has a third pole at 
the corotation point for disks with non-zero temperature gradients. This indicates
that corotation is an irregular singular point, making it challenging to find 
a 3D wave solution for disks with radial temperature gradients.

In this study, we attempt to obtain the Type I migration torque for 3D 
locally isothermal disks with radial temperature gradients, by suppressing 
the divergence at corotation in the linear wave equation with the Landau
prescription. Our linear calculation gives a definite torque on the planet
despite the {singularity} at corotation.
In the next section, we describe the formulation of our linear wave 
calculation with the Landau prescription for 3D locally isothermal disks. 
We also explain the origin of the strong divergence at corotation in 3D 
locally isothermal disks in the subsections 2.3 and 2.4. In Section 3, we 
present the wave solutions and the torques due to the waves. We will show that 
the obtained torques do not depend on the damping parameter of the Landau 
prescription. Then, we also present the linear torque formulas for 3D locally 
isothermal disks. 
In Section 4, we compare the linear torque formula with the results of 
the previous 3D hydro-dynamical simulations.
As well as the linear corotation torque, our linear calculation can 
evaluate the 3D horseshoe torque, which is also compared with the 
previous 3D simulations. In the last section, we summarize our findings.
\vspace{2mm}

\section{Formulation for locally isothermal disks}\label{sec:formulation}
\subsection{Model and Assumptions}\label{subsec:model}
We adopt the following assumptions for density waves in three-dimensional disks, 
as done by Paper I

%
\begin{itemize}
\item[1.] 
The amplitudes of the density waves excited by a planet are small enough for the 
linear calculation to be valid. We also assume that the mass of the planet is so
small that the disk gap does not form. The orbit of the planet is circular and 
lies in the midplane of the disk (i.e., $e=i=0$).
\item[2.]
We consider a locally isothermal disk that is vertically isothermal, but radially 
non-isothermal. We also assume an isothermal equation of state. The disk temperature 
and the surface density of the unperturbed disk are both power-law functions of the 
radial coordinate $r$. The self-gravity of the disk is negligibly small.

\item[3.]
The boundary condition is the non-reflecting one for density waves propagating away
from the planet. In this paper, the waves are damped near the corotation point with
the Landau prescription. Thus, the density waves propagating towards corotation are
not reflected at corotation (see Section 3). The Landau prescription is necessary 
to suppress the strong divergence at corotation.
\end{itemize}

\subsection{Basic Equations and the Unperturbed Disk}\label{subsec:basiceq}
The basic equations are the Euler equation and the equation of continuity given by
\begin{equation}
\frac{ \partial {\bf v} }{\partial t } + ({\bf v } \cdot \nabla ){\bf v}
= -\frac1\rho \nabla p - \nabla \left(- \frac{GM\sub{*} }{|{\bf r }| }\right)
  - \nabla \phi\sub{p},
\label{eulereq}
\end{equation}
\begin{equation}
\frac{ \partial \rho }{\partial t } + \nabla \cdot (\rho{\bf v })=0,
\label{eqofcont}
\end{equation}
where $\rho$, $p$, $\vect{v}$ are the density, the pressure, and the velocity
of the disk gas, respectively, and $M_*$ is the mass of the host star.
The gravitational potential of the planet with the mass $M_p$
is given by $\phi\sub{p} = -G M\sub{p}/|{\bf r} - {\bf r}\sub{p}|$,
where ${\bf r}\sub{p}$ is the position vector of the planet.
To describe the disk, we use a cylindrical coordinate system, ($r$, $\theta$, $z$).
The $z$-axis is chosen to coincide with the rotation axis of the disk.
The unperturbed surface density $\sigma$ and the temperature $T$ of the disk
depend on $r$ as\footnote{
The readers should note that the definition of $\beta$
of the present paper is different from that in Paper I by factor 2.
In Paper I, $\beta$ was the power-law index of the sound speed.}
\begin{equation}
\sigma = \sigma_p (r/r_p)^{-\alpha}, \qquad T = T_p (r/r_p)^{-\beta},
\end{equation}
where $r_p = |{\bf r}\sub{p}|$ is equal to the radius of the circular orbit
of the planet.
We adopt the isothermal equation of state $p = c(r)^2 \rho$.
Since the isothermal sound speed $c$ depends on $r$ as
$c=c_p(r/r_p)^{-\beta/2}$, the disk is baroclinic.

The unperturbed disk is steady and axisymmetric.
The vertical density distribution of the unperturbed disk is
\begin{equation}
\rho_0 = \frac{\sigma}{\sqrt{2\pi} h} \exp \left( -\frac{z^2}{2h^2}\right),
\label{rho0}
\end{equation}
where the disk scale height $h$ is given by $h = c/\Omega_K$ and 
$\Omega_K \equiv \sqrt{GM_*/r^3}$ is the Keplerian angular velocity on the midplane.
The disk scale height depends on $r$ as $h= h_p (r/r_p)^\mu$ and the power-law index 
is given by $\mu = 3/2 - \beta/2$. The small disk aspect ratio, $h/r$, is assumed in 
Equation~(\ref{rho0}).
On the disk midplane ($z=0$), the pressure also has the power-law radial distribution 
of $p \propto r^{-\delta}$, where the power-law index is $\delta = 3/2+\alpha+\beta/2$. 

The angular velocity $\Omega $ of the unperturbed disk rotation is
\begin{equation}
\Omega = \Omega\sub{K}
 \left\{ 1 - \frac12 \left(\frac{h_p}{r_p} \right)^{\!2}
\!\left[ \frac32 + \alpha
 + \frac{\beta}{2} \left( \frac{z^2}{h_p^2} +1 \right) \right] \right\},
\label{omega}
\end{equation}
where the terms of $O((h/r)^4)$ and the higher are neglected.
Note that $\Omega $ depends also on $z$ in radially non-isothermal disks.

\subsection{Fourier-Hermite Components of the Perturbation Equations
}\label{subsec:hermite}

We describe the equations for the velocity and pressure perturbations 
$\boldsymbol{u}$ and $p_1$. We also introduce a perturbation $\eta$ defined by
\begin{equation}
\eta = {p_1}/{\rho_0} = c^2{\rho_1}/{\rho_0}.
\label{eta_def}
\end{equation}
Although $\eta$ defined by Equation~(\ref{eta_def}) is the enthalpy perturbation 
in completely isothermal disks, it is not for the radially non-isothermal case.
We expand the perturbations with Fourier series and Hermite polynomials, as in
Paper I.
\begin{equation}
\eta(r, \theta, z) = \!
\sum^\infty _{m, n=0}
\! \Re \left[\,
\eta_{m,n}(r) \, H_n(Z)\, \mbox{e}^{im(\theta - \Omega\sub{p} t )+\epsilon t}
\,\right],
\label{eta2}
\end{equation}
where $Z$ is the non-dimensional vertical coordinate defined by $Z=z/h(r)$
and $\Re$ represents the real part.
The coefficients of the summation (e.g., $\eta_{m,n}(r)$) are called the 
Fourier-Hermite components. In this paper, we consider slowly growing 
perturbations, by adding the factor 
$\mbox{e}^{\epsilon t}$ in Equation~(\ref{eta2}) with a positive constant 
$\epsilon$, for the Landau prescription. Since $\epsilon$ is set to be much 
smaller than $\Omega_p$, the Landau prescription is effective only near 
corotation.

The linearized Euler equation and equation of continuity for the 
"Fourier components" are given by 
\begin{align}
& \left[ im(\Omega - \Omega\sub{p}) + \epsilon \right] u_{r,m} - 2 \Omega u_{\theta,m}
= - \frac{\partial \eta'_{m}}{\partial r} - \frac{\beta}{r}  \eta_{m},
\label{euler_r0}
\\[2mm]
& \left[ im(\Omega - \Omega\sub{p}) + \epsilon \right] u_{\theta,m} + 2 B u_{r,m}
= - \frac{im}{r}  \eta'_{m},
\label{euler_theta0}
\\[2mm]
& \left[ im(\Omega - \Omega\sub{p}) + \epsilon \right] u_{z,m} 
= - \frac{\partial \eta'_{m}}{\partial z},
\label{euler_z0}
\\[2mm]
& \left[ im(\Omega - \Omega\sub{p}) + \epsilon \right] \eta_{m} /c^2 
+ \frac{d u_{r,m}}{d r} + \frac{\partial \ln (r \rho_0)}{\partial r} u_{r,m}
\nonumber \\
& \qquad + \frac{im}{r} u_{\theta,m} 
+ \frac{\partial u_{z,m}}{\partial z} - \frac{z}{h^2} u_{z,m} =0,
\label{cont_0}
\end{align}
where, $\eta'_{m} = \eta_{m} + \phi_{p,m}$.
The terms of $\epsilon$ represent the Landau prescription.
These terms suppress the divergence of waves at corotation.

We further expand these equations with Hermite polynomials 
(\citealt{1998PASJ...50..141T}; Paper I).
For the expansion, each equation is multiplied by
$H_n(Z)\,\mbox{e}^{-Z^2/2}/(\sqrt{2 \pi} n!)$
and integrated with respect to $z$.
For example, the term, $im(\Omega - \Omega\sub{p}) u_{r,m}$, is transformed as
\begin{align}
&\int^\infty_{-\infty} im(\Omega - \Omega\sub{p}) u_{r,m}
\frac{1}{\sqrt{2 \pi} n!} H_n(Z) \mbox{e}^{-Z^2/2} dZ \nonumber \\
& = im ( \Omega_n - \Omega_p ) u_{r,m,n}\nonumber \\
& \quad + im \Delta \Omega [ (n+1)(n+2) u_{r,m,n+2} + u_{r,m,n-2} ],
\label{u_r_hermite}
\end{align}
where
\begin{align}
&\Delta \Omega = - \frac14 \left(\frac{h_p}{r_p}\right)^2 \beta \Omega_p,
\label{deltaomega} \\[2mm]
&\Omega_n 
= \Omega\sub{K}
\left[1-\frac12 \left(\frac{h_p}{r_p} \right)^{\!2} 
 \!\left(\frac32 + \alpha + \frac{\beta}{2}  \right) \right]
+(2n+1) \Delta \Omega,
\label{omegan}
\end{align}
and we used the normal orthogonal relation between polynomials $H_n$ 
given by 
\begin{equation}
\int^\infty_{-\infty} e^{-Z^2/2} H_m(Z)H_n(Z) dZ = \sqrt{2\pi}n!\delta_{m,n}.
\end{equation}
In Equation~(\ref{u_r_hermite}), noting that $|r-r_p|/h_p, mh_p/r_p \sim O(1)$,
we neglected the terms of O$((h_p/r_p)^3 )$ and the higher.
For radially non-isothermal disks,
$\Delta \Omega$ has a non-zero value and $\Omega_n$ depends on $n$.
Transforming Equations~(\ref{euler_r0}) and (\ref{cont_0}) in a similar way, 
we obtain
\begin{align}
&\frac{d \eta'_{m,n}}{d r}  = \nonumber \\
&\quad  -\left[ im(\Omega_n - \Omega\sub{p}) + \epsilon \right]
u_{r,m,n} +2 \Omega\sub{K} u_{\theta,m,n}  \nonumber \\
& \quad + \frac{\mu}{r} [n \eta'_{m,n} + (n+1)(n+2) \eta'_{m,n+2}]
- \frac{\beta}{r} \eta_{m,n}
\nonumber \\[1mm]
& \quad
- i m \Delta \Omega [ (n+1)(n+2) u_{r,m,n+2} + u_{r,m,n-2} ],
\label{eulereq_r}
\\[2mm]
&\frac{d u_{r,m,n}}{d r} = \nonumber \\
&\quad - \left[ im(\Omega_n - \Omega\sub{p}) + \epsilon \right] \eta_{m,n} /c^2
- \frac{im}{r} u_{\theta,m,n}  \nonumber \\
& \quad 
+ \frac{u_{z,m,n-1}}{h} -  \frac{1-\alpha}{r} u_{r,m,n}
-  \frac{\mu}{r} ( n u_{r,m,n} + u_{r,m,n-2})
\nonumber \\[1mm]
& 
\quad -i m \Delta \Omega [ (n+1)(n+2) \eta_{m,n+2} + \eta_{m,n-2}]/c^2,
\label{eqofcont2}
\end{align}
\vspace{2mm}
Equations (\ref{euler_theta0}) and (\ref{euler_z0}) are transformed as
\begin{align}
& [ m( \Omega_n - \Omega_p ) -i \epsilon ] u_{\theta,m,n} = 
- \frac{m}{r} \eta'_{m,n} + \frac{i}{2} \Omega\sub{K} u_{r,m,n}
\nonumber \\[1mm]
& \quad - m \Delta \Omega [ (n+1)(n+2) u_{\theta,m,n+2} + u_{\theta,m,n-2} ],
\label{eulereq_theta}
\\[3mm]
& [ m( \Omega_{n-1} - \Omega_p ) -i \epsilon ] u_{z,m,n-1 }=  
i \frac{n}{h} \eta'_{m,n} 
\nonumber \\[1mm]
& \quad - m \Delta \Omega [ n(n+1) u_{z,m,n+1} + u_{z,m,n-3} ].
\label{eulereq_z}
\end{align}
In Equations~(\ref{eulereq_r})-(\ref{eulereq_z}), the terms proportional to 
$\beta$ (or $\Delta \Omega$) come from the radial temperature gradient. Due 
to the non-zero $\Delta \Omega$, each equation has coupling terms with 
different $n$ modes. By eliminating the velocity perturbations $u_{i,m,n}$ 
in these equations, we can obtain a second-order differential equation for 
$\eta'_{m,n}$ (i.e., a wave equation). Due to these eliminations, strong 
divergences appear at corotation of radially non-isothermal disks.

When $\beta=0$, $\Delta \Omega$ vanishes and $\Omega_n$ is independent of 
$n$. From Equations~(\ref{eulereq_theta}) and (\ref{eulereq_z}), we see 
that the perturbations $u_{\theta,m,n}$ and $u_{z,m,n}$ are proportional to 
$1/(\Omega_n-\Omega_p)$ and diverge at corotation, where $\Omega_n=\Omega_p$.
Then, in the vicinity of corotation, the wave equation has a form
\begin{equation}
\frac{d^2}{dr^2}\eta'_{m,n} 
+ \frac{n \Omega_p^2}{h_p^2 m^2 (\Omega_n-\Omega_p)^2} \eta'_{m,n} =0
\end{equation}
when $\epsilon =0$.
Thus, the wave equation has the second-order pole at corotation where 
$\Omega_n=\Omega_p$, but it is a regular singular point.

For radially non-isothermal disks with a non-zero $\Delta \Omega$, 
the corotation points of each $n$ mode are located at slightly different radial 
positions. Furthermore, Equations~(\ref{eulereq_theta}) and (\ref{eulereq_z})
have the coupling terms with adjacent $n'$ modes, and the adjacent $n'$ mode
waves are also coupled with other modes. Because of this coupling chain, 
each $n$ mode wave is influenced by the corotation resonances of all the other 
modes. This coupling effect makes the divergence at corotation stronger than 
that in a completely isothermal disk. In the next subsection, we will find that 
the wave equation has the third-order pole at corotation under the modified 
local approximation, as also shown in Paper I. 
Note that the terms proportional to $im(\Omega_n - \Omega_p )$ in 
Equations~(\ref{eulereq_r}) and (\ref{eqofcont2}) do not cause any divergence. 
Thus, we will neglect the terms with $\epsilon$ in Equations~(\ref{eulereq_r}) 
and (\ref{eqofcont2}) below.

\subsection{Modified local approximation}
Furthermore, the perturbations and their equations are expanded
with the small disk aspect ratio, $h_p/r_p$.
We define the local radial coordinate $x$ as\footnote{
From this definition of $x$, we obtain the position of the planet
in the $x$-coordinate as $x_p = (\delta/3) h_p/r_p$.
Since this expression of $x_p$ is the same as that in Paper I,
the expression of $\Phi_{m,n,1}$ is also the same as that in Paper I.}
\begin{equation}
x = (r-r\sub{c,mid})/h_p,
\label{x}
\end{equation}
where the corotation radius at the midplane, $r\sub{c,mid}$, is defined by
$\Omega(r\!=\!r\sub{c,mid},z\!=\!0)=\Omega_p$. By using $x$, the augular velociy 
$\Omega_n$ defined by Equation~(\ref{omegan}) is given by
\begin{equation}
\Omega_n = \left(1 + \frac{3 h_p}{2r_p} x 
+ \frac{15 h_p^2}{8 r_p^2} x^2 \right)
\Omega_p + 2(n+1)\Delta \Omega.
\end{equation}
The unperturbed surface density, sound velocity, and scale height are expanded as
$\sigma= (1 -\alpha \frac{h_p}{r_p} x) \sigma_p$, 
$c= (1- \frac{\beta}{2} \frac{h_p}{r_p}x)c_p$, 
and $h= (1+\mu  \frac{h_p}{r_p} x)h_p$, respectively.
The perturbation components are expanded as
\begin{align}
\phi_{p,m,n} =& \lambda c_p^2 
\left( \Phi_{m,n,0} + \frac{h_p}{r_p} \, \Phi_{m,n,1} \right), 
\nonumber \\[2mm]
\eta_{m,n} =& \lambda c_p^2 
\left( W_{m,n,0} + \frac{h_p}{r_p} \, W_{m,n,1} \right), 
\nonumber \\[2mm]
u_{i,m,n} =& \lambda c_p 
\left( U_{i,m,n,0} + \frac{h_p}{r_p} \, U_{i,m,n,1} \right), 
%
\label{mlocal}
\end{align}
where $\lambda = (M_p/M_*)(r_p\Omega_p/c_p)^2$ and the subscript $i$
indicates $r$, $\theta$, or $z$.
The expressions of the potentials $\Phi_{m,n,j}$ ($j=0$, 1) have been given 
by the Equation~(27) of Paper I. We also define 
$W'_{m,n,j} \equiv W_{m,n,j} +\Phi_{m,n,j}$, which is related to $\eta'_{m,n}$. 
The zeroth order perturbation $W_{m,n,0}$ corresponds to the shearing sheet 
model and the first order perturbation $W_{m,n,1}$ represents the deviation 
from the shearing sheet model, which gives the net torque on the planet.

We write down the equations for zeroth and first-order perturbations. 
From the zeroth and first-order terms with respect to $h_p/r_p$ in 
Equations~(\ref{eulereq_r}) to (\ref{eulereq_z}), we obtain
\begin{align}
&\frac{d W'_{m,n,j}}{dx} = i\frac32 k_\theta x U_{r,m,n,j} 
+2 U_{\theta,m,n,j} + Q_j^{(W')}, \label{eulereq_r_local}
\\[2mm]
&\frac{d U_{r,m,n,j}}{dx} = i\frac32 k_\theta x W_{m,n,j}
- i k_\theta U_{\theta,m,n,j} \nonumber \\
& \qquad \qquad \quad + U_{z,m,n-1,j}
+Q_j^{(U_r)},
\label{eqofcont_local}
\\
%
%
&U_{\theta,m,n,j} =
\left( -\frac32 k_\theta x - i \varepsilon \right)^{\!\!-1} \nonumber \\
& \qquad \qquad \quad 
\times \left(-k_\theta W'_{m,n,j} + \frac{i}{2} U_{r,m,n,j} + Q_j^{(U_\theta)}
\right),
\\[2mm]
&U_{z,m,n-1,j} =
\left( -\frac32 k_\theta x - i \varepsilon \right)^{\! -1} \!
\left( i n W'_{m,n,j} + Q_j^{(U_z)} \right),
\label{eulereq_z_local}
\end{align}
where $k_\theta = m h_p/r_p$, $\varepsilon = \epsilon/\Omega_p$ and $j=0$ or 1 
in each equation. 
The equation with $j=0$ is that for the 
shearing sheet model and $j=1$ corresponds to the equation for the 
first-order perturbation. In Equations~(\ref{eulereq_r_local}) and
(\ref{eqofcont_local}),
terms with $\varepsilon$ are omitted.
Moreover, the source terms $Q_0^{(X)} =0$ and $Q_1^{(X)}$'s are given by
\footnote{
The potentials $\Phi_{m,n,j}$ are also source terms. Those are included 
in Equation~(\ref{eqofcont_local}) through 
$W_{m,n,j} = W'_{m,n,j} -\Phi_{m,n,j}$.}.
\begin{align}
Q_1^{(W')} =& 
- i\frac{15}{8} k_\theta x^2 U_{r,m,n,0} -3x U_{\theta,m,n,0}
- \beta W_{m,n,0}
\nonumber\\
& 
+ \mu \left[ (n+1)(n+2)W'_{m,n+2,0} + n W'_{m,n,0} \right]
\nonumber\\
& 
+ i \beta \, \frac{k_\theta}{4}
\left[ (n+1)(n+2) U_{r,m,n+2,0} \right.
\nonumber\\
& \qquad \quad \: \:
\left. +(2n+1)U_{r,m,n,0} 
+ U_{r,m,n-2,0} \right],
%
\end{align}
\begin{align}
Q_1^{(U_r)} =& 
- i \left(\frac{15}{8} - \frac32\beta \right) k_\theta x^2 W_{m,n,0}
+ i k_\theta x U_{\theta,m,n,0}
\nonumber\\
& 
- (1-\alpha)  U_{r,m,n,0} - \mu x U_{z,m,n-1,0}
\nonumber\\
& 
- \mu \left[ n U_{r,m,n,0}+U_{r,m,n-2,0} \right]
\nonumber
\\
& 
+ i \beta \, \frac{k_\theta}{4}
\left[ (n+1)(n+2) W_{m,n+2,0} \right.
\nonumber\\
& 
\left. \qquad \quad \: \:
+(2n+1)W_{m,n,0} 
+ W_{m,n-2,0} \right],
\end{align}
\begin{align}
Q_1^{(U_\theta)} =& - \frac{15}{8} k_\theta x^2 U_{\theta,m,n,0}
+ k_\theta x W'_{m,n,0}
- i \frac34 x U_{r,m,n,0} 
\nonumber\\
& 
+ \beta \, \frac{k_\theta}{4}
\left[ (n+1)(n+2) U_{\theta,m,n+2,0} \right.
\nonumber\\
& \left. \qquad \quad \: \:
+(2n+1)U_{\theta,m,n,0} + U_{\theta,m,n-2,0} \right],\\
%
Q_1^{(U_z)} =& -\frac{15}{8} k_\theta x^2 U_{z,m,n-1,0}
- i \mu n x W'_{m,n,0} 
\nonumber\\
& 
+ \beta \, \frac{k_\theta}{4}
\left[ n(n+1) U_{z,m,n+1,0} \right.
\nonumber\\
& \left. \qquad \quad \: \:
+(2n-1)U_{z,m,n-1,0} + U_{z,m,n-3,0} \right].
\label{Q_1z}
\end{align}
In the above definitions, $Q_1^{(X)}$ consist of terms proportional to
the temperature gradient ($\beta$), surface density gradient ($\alpha$), or
scale height variation ($\mu$), and other terms (that are called 
"curvature terms"). The terms proportional to the pressure gradient ($\delta$) 
appear in the potential $\Phi_{m,n,1}$ (see Paper I). 
Since Equations~(\ref{eulereq_r_local}) to (\ref{eulereq_z_local}) are linear,
the first-order solution is written by the superposition of terms related 
to each effect.
\begin{align}
W'_{m,n,1} = & W'^{(\mbox{\scriptsize curv})}_{m,n,1}
+\alpha W'^{(\alpha)}_{m,n,1}
+\beta W'^{(\beta)}_{m,n,1} \nonumber \\
& +\delta W'^{(\delta)}_{m,n,1}
 +\mu W'^{(\mu)}_{m,n,1}.
\label{solution_sum}
\end{align}
The main target of this paper is the term related to the temperature gradient, 
$W'^{(\beta)}_{m,n,1}$ since other terms have obtained in Paper I.

We check the divergence at corotation again for
Equations~(\ref{eulereq_r_local})-(\ref{eulereq_z_local}) with $\varepsilon=0$. 
In the zeroth order equations, $U_{\theta,m,n,0}$ and $U_{z,m,n,0}$ have the 
first-order pole at corotation and the wave equations with $j=0$ have the 
following form in the vicinity of corotation.
\begin{equation}
\frac{d^2}{dx^2}W'_{m,n,0} +\frac{2i}{3k_\theta x} U_{z,m,n+1,0}=0.
\end{equation}
It has the second-order pole but it is a regular singular point.

For the first-order equations, we have the same wave equation but $j=1$.
In this case, $U_{z,m,n+1,1}$ has the source term $Q^{(U_z)}_1$, which
includes the terms $U_{z,m,n\pm1,0}$ and $U_{z,m,n-3,0}$ for a non-zero 
$\beta$ (see Equation~[\ref{Q_1z}]). Since these terms are proportional to 
$1/x$, $U_{z,m,n+1,1}$ is proportional to $1/x^2$ and the wave equation has 
the third pole at corotation, which { causes a non-removal singular point
in the wave solution.} In this paper, we solve the wave equation, by 
suppressing the strong divergence at corotation with the Landau prescription.

Note that {the term with the third pole} at $x=0$ is caused by the modified 
local approximation. In Equations~(\ref{eulereq_r}) to (\ref{eulereq_z}), 
the corotation points of each $n$ mode split into slightly different 
radial positions due to the $n$-dependence of $\Omega_n$. 
In Equations~(\ref{eulereq_r_local})-(\ref{eulereq_z_local}), on the other 
hand, the corotation points of all $n$ modes degenerate to $x=0$. This 
degeneracy of corotation leads to {the third pole}. In the 
next section, we will examine how this strong divergence at corotation 
affects the density waves and the planetary torques.

\subsection{Boundary Conditions}
The radiation boundary condition is applied to the solutions at 
large $|x|$, as done in Paper I. The radiation condition is written by
\begin{equation}
\frac{d\eta'_{m,n}}{dr} = i s(r) \eta'_{m,n}.
\label{bound_con_inf}
\end{equation}
In the modified local approximation, $s(r)$ is expanded as
\begin{equation}
s= [S_0(x) + (h_p/r_p) S_1(x)]/h_p,
\end{equation}
and $S_j(x)$'s are given by
\begin{eqnarray}
S_0&=& \displaystyle
\frac32 k_\theta |x| - \frac{k_\theta^2 + n + 1}{3k_\theta |x|}
\nonumber\\
&& \displaystyle + \, \frac{51+ 16n -4(k_\theta^2+n+1)^2}{108 k_\theta^3 |x|^3}
\nonumber\\
&&\displaystyle +\,i \left(
-\frac{1}{2x}
+2\frac{ k_\theta^2 + n -1 }{ 9 k_\theta^2 x^3 }
\right), \\[3mm]
S_1&=&\displaystyle
x \left[\, 
\left( -\frac{15}{8} + \frac32 \beta \right)
k_\theta |x| \right.
\nonumber\\
&&\displaystyle \left. \quad \:
+ \frac{7+3k_\theta^2-5n +8n \mu +4(k_\theta^2-1)\beta
}{12k_\theta |x|}
\right]
\nonumber\\
&&\displaystyle
+\,i \left( \frac98 - \frac12 \alpha +\frac32 \beta \right).
\end{eqnarray}
Equation~(\ref{bound_con_inf}) gives the boundary conditions for 
$W'_{m,n,j}$ using $S_0$ and $S_1$. To seek the solution related to 
the temperature gradient $W'^{(\beta)}_{m,n,1}$, we leave only terms 
proportional to $\beta$ in $S_1$ and put $\beta=1$. The other terms 
in $S_1$ are omitted.

As in Paper I, the general solutions of $W'_{m,n,j}$ are written as
$W'_{m,n,j} = A_{1,j} y_1 + A_{2,j} y_2 + W'_{m,n,j,p}$,
where $y_1$ and $y_2$ are two independent solutions to the homogeneous 
differential equations~(\ref{eulereq_r_local}) and (\ref{eqofcont_local}),
and $W'_{m,n,j,p}$ is a {particular} solution to the inhomogeneous equations.
The complex coefficients $A_{i,j}$ are determined by the boundary 
conditions at $x = \pm \infty$.
We adopt the following boundary conditions at $x=0$.
\begin{align}
&y_1=1, \qquad y_2 =W'_{m,n,j,p}=0, \nonumber \\
&\frac{dy_1}{dx} = \frac{dW'_{m,n,j,p}}{dx} =0, 
\qquad \frac{dy_2}{dx} = 1, \qquad \mbox{($x=0$)}.
\label{bcond_W'_at_0}
\end{align}
Noting that the solutions $W'_{m,n,j}$ have symmetries as
\begin{align}
W'_{m,n,0}(-x)&=W'^*_{m,n,0}(x), \nonumber \\[1mm]
W'_{m,n,1}(-x)&=-W'^*_{m,n,1}(x)
\label{symmetries_W'}
\end{align}
(see Paper I),
we find that the coefficients $A_{1,0}$ and $A_{2,1}$ are real and 
that $A_{2,0}$ and $A_{1,1}$ are purely imaginary numbers. 
Therefore, we only have to integrate the equations~(\ref{eulereq_r_local}) and 
(\ref{eqofcont_local}) from $x=0$ to $+\infty$.

\subsection{Total and Corotation Torques}\label{subsec:torque formulation}

A torque exerted by the planet on a ring region with a radius $r$ 
and a width of the unit length of the disk
(i.e., the torque density) is given by
\begin{equation}
\frac{d\Gamma}{dr} = - \iint dz \, d\theta \, r \frac{\rho_0 \eta}{c^2}  
\frac{\partial \phi_p}{\partial \theta},
\label{torqden0}
\end{equation}
where the relation $\rho_1 = \eta \rho_0/c^2$ is used.
By integrating Equation~(\ref{torqden0}), we have the torque on the 
entire disk, and the opposite sign is the torque on the planet.
The torque density can be written as the sum of the contributions
from each Fourier-Hermite component.
\begin{equation}
\frac{d\Gamma}{dr} = \sum_{m,n} \frac{d\Gamma_{m,n}}{dr}.
\end{equation}
Since $\eta$ and $\phi_p$ are expanded as Equation~(\ref{eta2}),
the contributions from each Fourier-Hermite component 
are given by
\begin{equation}
\frac{d\Gamma_{m,n}}{dr} = - \pi n ! m \frac{r \sigma}{c^2}
\Im(\eta'_{m,n}) \phi_{p,m,n},
\end{equation}
where $\Im$ represents the imaginary part.
Moreover, the expansions in the modified local approximation of 
Equations~(\ref{x}) and (\ref{mlocal}) yield
\begin{equation}
\frac{d\Gamma_{m,n}}{dr} = \frac{\Gamma_0}{h_p} 
\left[
\frac{d}{dx}T_{m,n,0} + \frac{h_p}{r_p} \frac{d}{dx}T_{m,n,1}
\right],
\label{dGmndr}
\end{equation}
where the zeroth- and first-order integrated torques $T_{m,n,j}$ are given by
\begin{align}
& T_{m,n,0}(x)=
- \pi n! \, k_\theta
\int^x_{-\infty} \Im(W'_{m,n,0}) \Phi_{m,n,0} \, dx,
\\
& T_{m,n,1}(x)= \nonumber \\
& \displaystyle
- \! \pi n! \, k_\theta \!
\int^x_{-\infty} \!
\left[
\Im(W'_{m,n,1}) \Phi_{m,n,0}+\Im(W'_{m,n,0}) \Phi_{m,n,1} \right] dx
\notag\\
&\displaystyle
+ (1-\alpha+\beta) \int^x_{-\infty} x \frac{d}{dx} T_{m,n,0} \, dx
\end{align}
and
\begin{equation}
\Gamma_0 = \left( \frac{M_p}{M_c} \frac{r_p}{h_p}\right)^2 
\sigma_p r_p^4 \Omega_p^2.
\end{equation}
The definition of $\Gamma_0$ is different from that in Paper I.
Note that $T_{m,n,0}(\infty)=0$ due to the symmetry of the sheering sheet 
model. Then, integrating Equation~(\ref{dGmndr}), we find that
$\Gamma_{m,n} = \Gamma_0 (h_p/R_p)T_{m,n,1}(\infty) $.
The radial integration of the torque density gives the total torque on the
entire disk $\Gamma\sub{total}$. 
From Eq.~(\ref{dGmndr}), we obtain
\begin{align}
\Gamma\sub{total} &= \sum_{m,n}  \int^\infty_0 
\frac{d\Gamma_{m,n}}{dr} dr 
= \sum_{m,n} \Gamma_{m,n} = \sum_{n} \Gamma_T(n) 
\end{align}
with
\begin{equation}
\Gamma_T(n) \equiv \sum_m \Gamma_{m,n}
=\Gamma_0 \int^\infty_0 T_{m,n,1}(\infty) d k_\theta, 
\end{equation}
where we used 
$\textstyle \sum\limits_m = (r_p/h_p) \int dk_\theta$.

In Paper I, the torques are obtained from the angular momentum fluxes
of each wave component. However, the angular momentum fluxes vary due to 
the radial temperature gradient as well as due to the torque exerted by the planet.
For radially non-isothermal disks, therefore, the torques should be 
calculated using the above equations. 

Similar to the first-order solution $W'_{m,n,1}$ in 
Equation~(\ref{solution_sum}), the torque contributions are
written as the sum of the terms due to each effect.
\begin{align}
&\Gamma_{m,n} =  
\Gamma_{m,n}\superscript{(curv)}
+ \alpha \Gamma_{m,n}^{(\alpha)}
+ \beta \Gamma_{m,n}^{(\beta)}
+ \delta \Gamma_{m,n}^{(\delta)}
+ \mu \Gamma_{m,n}^{(\mu)},\nonumber \\
&\Gamma_T(n) = 
\Gamma\superscript{(curv)}_T(n)
+ \alpha \Gamma^{(\alpha)}_T(n)
+ \beta \Gamma^{(\beta)}_T(n) \nonumber \\
& \qquad \qquad + \delta \Gamma^{(\delta)}_T(n) + \mu \Gamma^{(\mu)}_T(n).
\label{Gamma_mn_sum}
\end{align}
In this paper, we will obtain the terms $\Gamma_{m,n}^{(\beta)}$, 
$\Gamma^{(\beta)}_T(n)$. The other terms have been obtained in Paper I.

Furthermore, taking the summation for $n$
and noting $\delta = \frac32 + \alpha + \beta/2$ and
$\mu = \frac32 - \beta/2$,
we obtain the total torque on the whole disk, $\Gamma\sub{total}$, as
\begin{equation}
\Gamma\sub{total} = \Gamma\sub{total}^{\alpha=\beta=0}
+ \alpha \frac{d\,  \Gamma\sub{total}}{d\alpha}
+ \beta \frac{d\,  \Gamma\sub{total}}{d\beta},
\end{equation}
where the coefficients are given by
\begin{align}
& \Gamma\sub{total}^{\alpha=\beta=0} = 
{\textstyle \sum\limits_n} \, 
\Gamma_T\superscript{(curv)}(n)
+\frac32 {\textstyle \sum\limits_n} \, \Gamma_T^{(\delta)}(n)
+\frac32 {\textstyle \sum\limits_n} \, \Gamma_T^{(\mu)}(n),
\\[1mm]
&
\frac{d\,  \Gamma\sub{total}}{d\alpha} = 
{\textstyle \sum\limits_n} \,  \Gamma_T^{(\alpha)}(n) 
+ {\textstyle \sum\limits_n} \, \Gamma_T^{(\delta)}(n),
\\[2mm]
&
\frac{d\,  \Gamma\sub{total}}{d\beta} = 
{\textstyle \sum\limits_n} \,  \Gamma^{(\beta)}_T(n) 
+ \frac12 {\textstyle \sum\limits_n} \,  \Gamma_T^{(\delta)}(n)
- \frac12 {\textstyle \sum\limits_n} \,  \Gamma_T^{(\mu)}(n).
\end{align}
The corotation torque is also expressed in a similar form. We can also 
obtain the Lindblad torque by substructing the corotation torque from 
the total torque.

\begin{figure*}[ht]
\includegraphics[height=11.5cm]{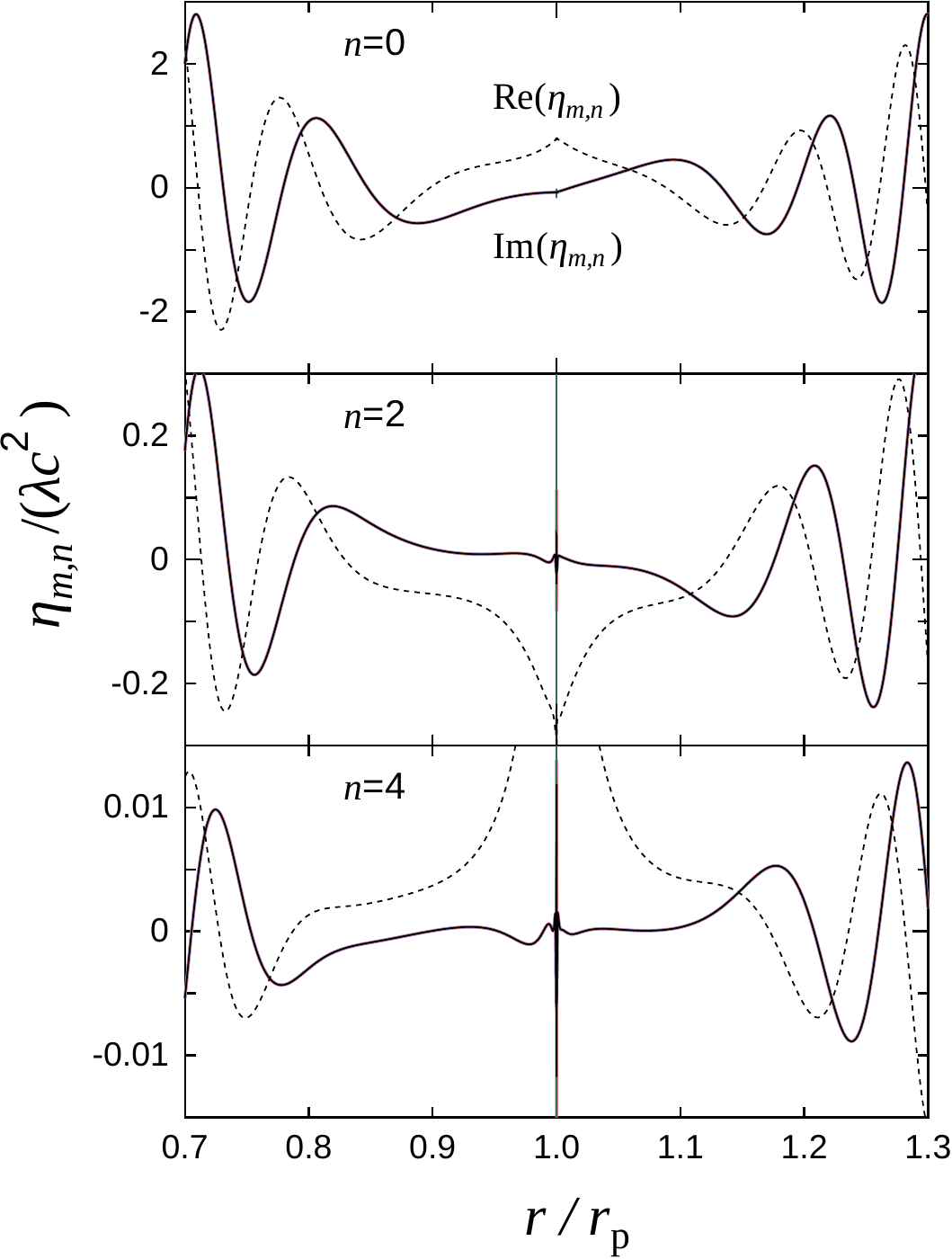}
\hspace{1mm}
\includegraphics[height=11.5cm]{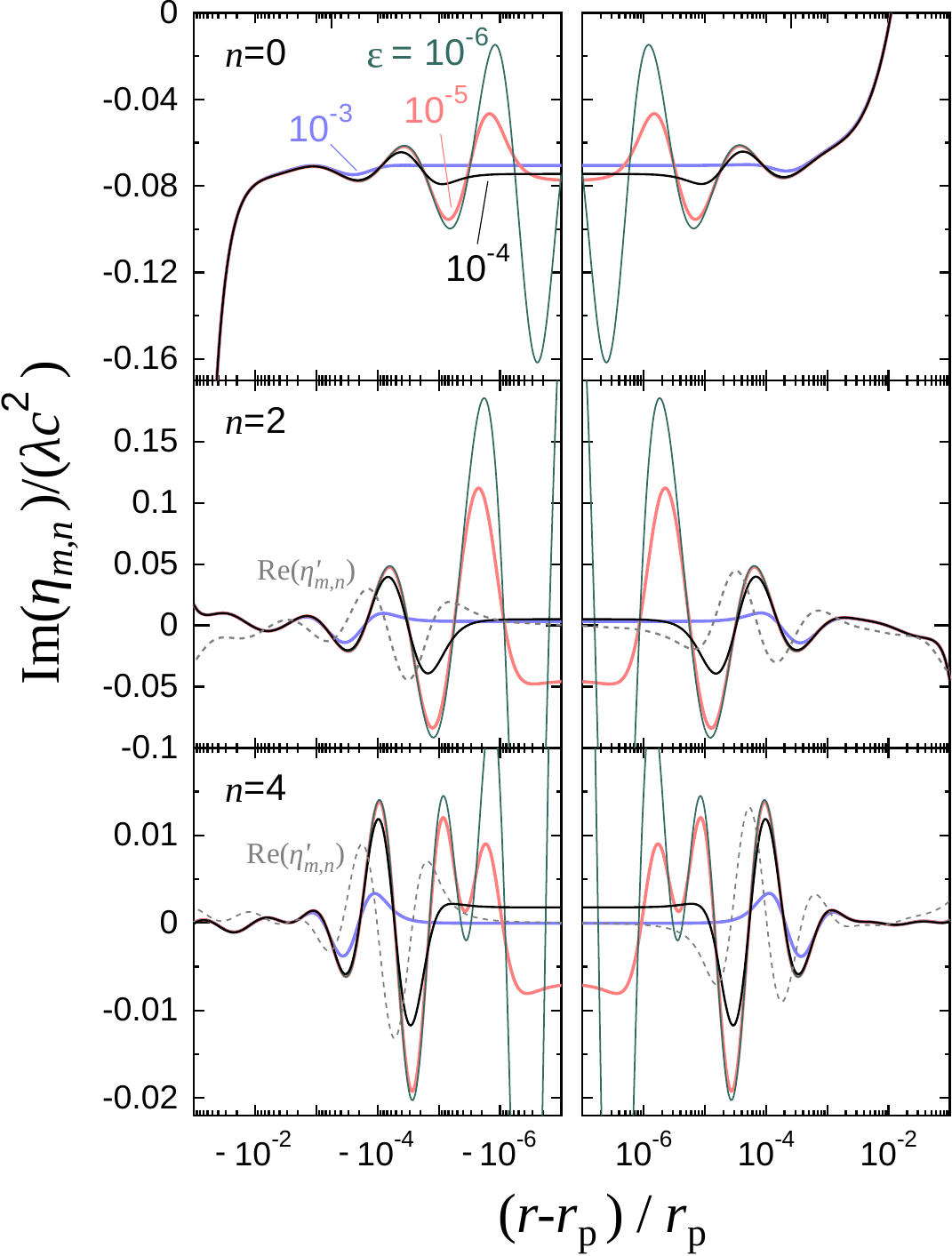}
\caption{
Waves in a radially non-isothermal disk. Fourier-Hermite components 
of the enthalpy perturbation are shown for $n=0$ (top), 2 (middle), 
and 4 (bottom). The left panel shows the full view and the right 
panel shows the magnified view near corotation. We set $\alpha=\mu=\delta=0$ 
and omit curvature source terms in the wave equations to focus on 
the effect of the radial temperature gradient. Other parameters are 
$\beta=2$, $m=10$, and $h_p/r_p=0.05$.
The black solid lines and dotted lines show the imaginary and real parts
of the wave with $\varepsilon=10^{-4}$. 
In the left middle and left bottom panels, the real parts of 
$\eta'_{m,n} (=\eta_{m,n}+\phi_{m,n})$ with $\varepsilon=10^{-4}$ are also
plotted with the gray dotted lines.
The imaginary parts of the solutions with $\varepsilon=10^{-3}$, $10^{-5}$, and $10^{-6}$
are plotted with the blue, red, {green} lines, respectively.
}
\label{fig:1}
\end{figure*}
We also derive the formula of the corotation torques due to two-dimensional 
waves ($n$=0) for the case of $\beta \ne 0$. Near corotation in radially 
non-isothermal disks, the wave equation of $\eta'_{m,0}$ is given by
\begin{equation}
\frac{d^2 \eta'_{m,0}}{dr^2} + \frac{P'}{r_c(r-r_c)} = 0
\end{equation}
with
\begin{align}
P' = \left\{
\frac{2\Omega}{d\Omega/dr}\right. &
\left[ \frac{d\ln(\sigma T/B)}{dr} \eta'_{m,0} \right. 
\nonumber \\
&	\; \; \left. \left. -\frac{d\ln T}{dr} \left( 
\phi_{m,0}
 - \frac{B}{\Omega} \phi_{m,2} 
\right) \right] \right\}_{r_c},
\end{align}
where we have used $\eta'_{m,2} \simeq 0$ at $r=r_c$ 
(e.g., Appendix of Paper I). Then, evaluating the jump in the angular 
momentum flux, we obtain the corotation torque by two-dimensional waves, 
$\Gamma\sub{C,$m$,0}$, as
\begin{align}
\Gamma\sub{C,$m$,0} &= 
\frac{\pi^2 m}{2}
\left\{
\frac{\sigma \eta'_{m,0}}{B d\Omega/dr}
\left[ \frac{d\ln(\sigma T/B)}{dr} \eta'_{m,0} \right. \right. \nonumber \\
& \hspace{26mm}\left. \left. 
	-\frac{d\ln T}{dr} \left( \phi_{m,0} - \frac{B}{\Omega} \phi_{m,2} 
\right) \right]
\right\}_{r_c}
\nonumber\\[1mm]
&= \frac{4\pi^2 k_\theta}{3} \left\{
\left[
\left(\alpha-\frac32 \right) W'_{m,0,0} +\beta W_{m,0,0}
\right. \right. \nonumber \\
& \hspace{17mm}\left. \left.
+ \frac{\beta}{4} \Phi_{m,2,0} 
\right]
W'_{m,0,0} \right\}_{x=0} 
\Gamma_0 \frac{h_p}{r_p}.
\label{ctorque}
\end{align}
When $\beta=0$, the above equation reduces to the expression in Paper I
(Eq.~[55]), in which isothermal disks are considered. 
The corotation torque is also written as the sum of terms due to each 
effect as in Equaition~(\ref{Gamma_mn_sum}).
The term due to the temperature gradient is given by
\begin{equation}
\Gamma\sub{C,$m$,0}^{(\beta)} = 
\frac{\pi^2 k_\theta}{3} (4 W_{m,0,0}+\Phi_{m,2,0})
W'_{m,0,0} \Gamma_0 \frac{h_p}{r_p}.
\end{equation}
In Paper I, the corotation torque also has contributions from 3D waves. 
However, as shown in the next section and Appendix~A, these torque 
contributions come from the 3D waves excited by the inner and outer 
near-side Lindblad resonances and should be included in the Lindblad torque. 
Therefore, in the present paper, the corotation torque is considered to 
result only from the 2D waves and is given by equation~(\ref{ctorque}).
\vspace{2mm}

\section{Numerical Results}

\subsection{Wave Solutions}
We obtain the Fourier-Hermite components of the enthalpy perturbation, 
$\eta_{m,n}$,
by integrating Equations~(\ref{eulereq_r_local}) and (\ref{eqofcont_local}) 
from $x=0$ to $+\infty$ with boundary conditions (\ref{bcond_W'_at_0}) 
and (\ref{bound_con_inf}), and using Equation~(\ref{mlocal}).
Figure~\ref{fig:1} shows typical solutions of $\eta_{m,n}$. 
Focusing on the effect of the disk temperature gradient, we set the 
first-order solution to be $W_{m,n,1} = \beta W_{m,n,1}^{(\beta)}$.
We can the solution $W_{m,n,1}^{(\beta)}$, by leaving only the 
source terms proportional to $\beta$ in the 
first-order equations~(\ref{eulereq_r_local}) to (\ref{eulereq_z_local}).
Using this first-order solution, we plot $\eta_{m,n}$ for $n=0, 2, 4$ 
in Figure~\ref{fig:1}. 
To emphasize the effect of the temperature gradient, we set $\beta=2$.
The left panel shows the full view and the right panel is the magnified 
view near corotation. Both the real and imaginary parts are plotted for 
solutions with the damping parameter $\varepsilon=10^{-4}$. Slight 
asymmetries of the waves for $r=r_p$ are due to the temperature gradient. 
The imaginary parts of the solutions with $\varepsilon=10^{-3}$, $10^{-5}$, and $10^{-6}$ 
are also plotted. The solutions with different $\varepsilon$ overlap almost 
completely, except near corotation.

\begin{figure*}[ht]
\includegraphics[height=12.6cm]{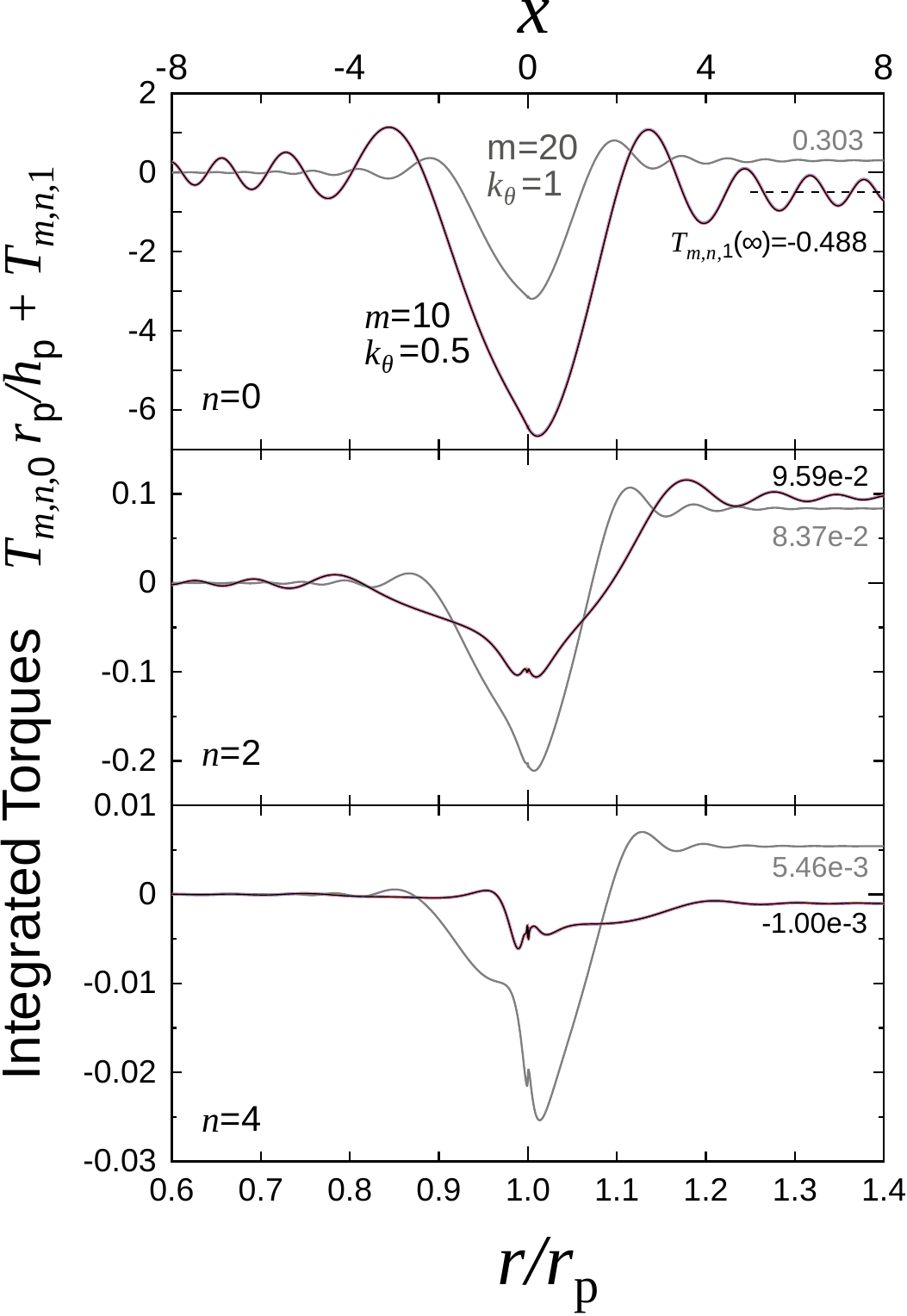}
\includegraphics[height=12.6cm]{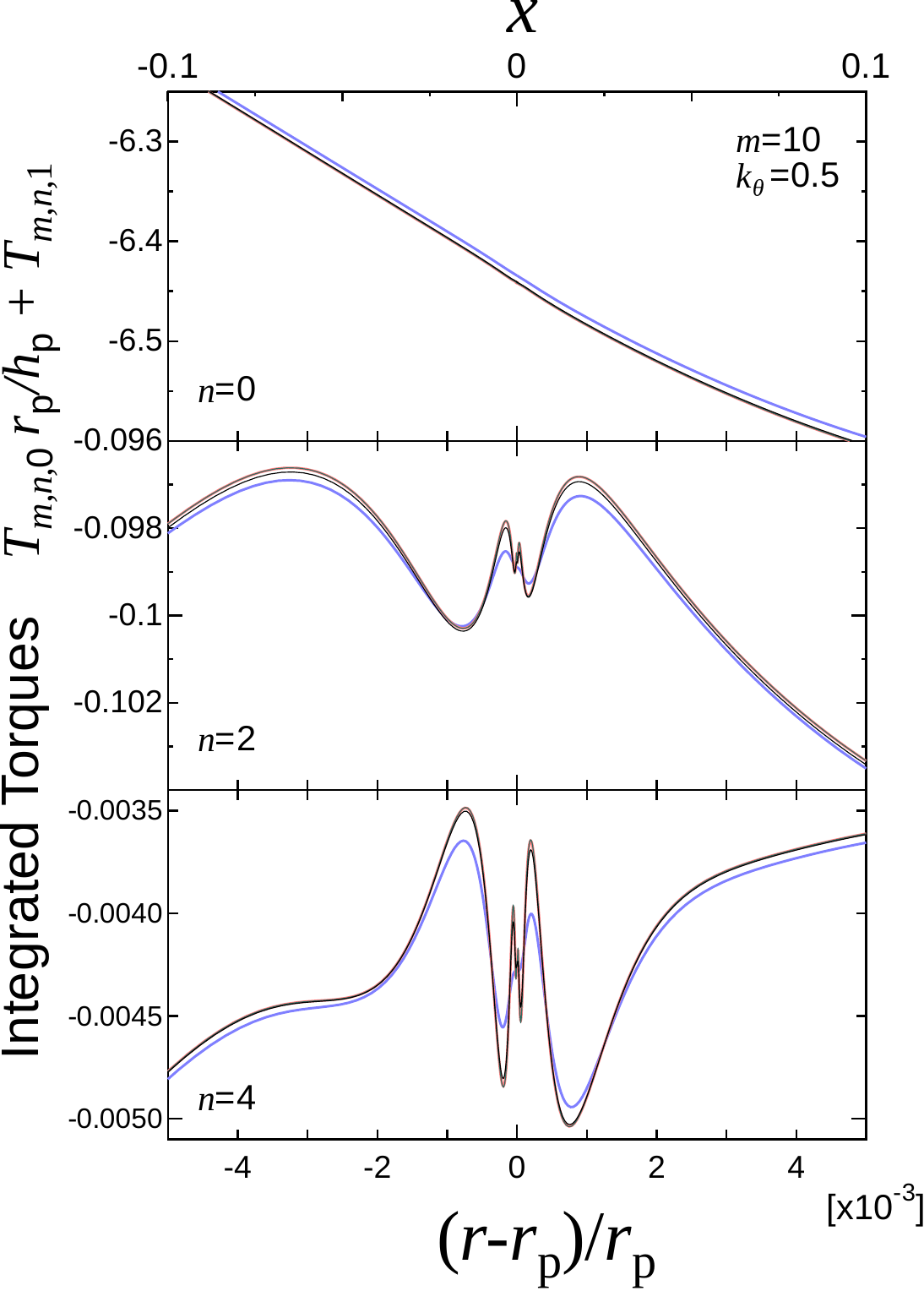}
\caption{The radial cumulative distribution of 
the torques due to the density waves.
The normalized integrated torques, 
$T_{m,n}/\Gamma_0 = T_{m,n,0} + (h_p/r_p)T_{m,n,1}$,
are plotted for the waves with $m=10$ (black lines), 20 (gray lines) 
and $n=0, 2, 4$ using $\varepsilon =10^{-4}$.
We also plot the integrated torque with $m=10$ for
$\varepsilon =10^{-3}$(blue lines), $10^{-5}$(red lines), and $10^{-6}$(green lines).
The left panel shows the full view and the right 
panel shows the magnified view near corotation.
}
\label{fig:2}
\end{figure*}

The wave solutions with the smaller $\varepsilon$ have stronger 
oscillations near corotation because a small $\varepsilon$ has
a weak damping effect. These oscillations result from the 
strongly divergent term in the wave equation for three-dimensional 
waves. Although the wave with $n=0$ also has some oscillations near 
corotation, the amplitude near corotation is much smaller than that 
of the waves propagating away from the planet (in the right panel). 
The oscillations of the wave with $n=0$ are caused by the interaction 
with the three-dimensional wave with $n=2$. 

Note that the strongly divergent term affects the wave only near 
corotation. The limited effect of the divergent term can be explained 
by the wave propagation. The excitation and propagation of density waves are 
described in Appendix A. The three-dimensional waves with 
$n \ge 2$ are excited at four Lindblad resonances for each $n$ mode 
(see Figure~\ref{figA}). The waves around the planet originate from the 
near-side Lindblad resonances and propagate towards corotation\footnote{
In the discussion of wave propagation in 
Appendix A, the trailing waves are expected. We find in 
Figure~\ref{figA} that the phase of the imaginary part of each wave 
always lags behind the real part, indicating that the waves are 
trailing. This is a natural consequence since we used the radiation 
boundary condition and the Landau prescription.
}.
These waves are expected to be dissipated at corotation due to their 
extremely short wavelength. Because of the direction of wave propagation, 
the divergent term at corotation has no effect outside the vicinity 
of corotation. 

In Paper I, we considered the torque associated with the waves propagating away 
from the planet to be the Lindblad torque and classified the torque associated with 
the three-dimensional waves propagating towards corotation as the corotation torque. 
However, the latter torque should be classified as the near-side Lindblad torque, 
since these waves are excited at the near-side Lindblad resonances.
Thus we adopt this new classification and consider the corotation torque to result 
only from the two-dimensional waves.

\subsection{Torques}
Next, we examine the torques exerted on the disk by the wave excitation.
Figure 2 shows the cumulative radial distributions of the torques 
due to the density waves in Figure~\ref{fig:1}.
The cumulative distribution of the torque due to $\eta_{m,n}$ is given by

\begin{equation}
\mathcal{T}_{m,n} \equiv
\int^r_0 \frac{d\Gamma_{m,n}}{dr} dr 
= \left( T_{m,n,0} + \frac{h_p}{r_p} T_{m,n,1} \right) \Gamma_0.
\end{equation}
We plot the cases where $\varepsilon = 10^{-3}, 10^{-4}$, $10^{-5}$, and $10^{-6}$,
The lines for each $\varepsilon$ almost overlap and show only small 
differences between them even near corotation (see the right panel).
Although there are some oscillations near corotation,
the amplitudes of the torque oscillations are small compared to
those of $\eta_{m,n}$ in Figure~\ref{fig:1}. Furthermore, the 
$\varepsilon$ dependence of the integrated torques is similar between 
before and after corotation, indicating that the contributions near the 
corotation are quite small for the integrated torques. 

By integrating the torque densities to a sufficiently large $r_p$, we 
obtain the convergence values of the torques due to each wave. Since 
the integrated torque converges slowly as it oscillates, we obtained the 
convergence value by calculating the average value of the oscillations. 
The convergence values of each integrated torque are shown in the 
left panel. From the convergence value $T_{m,n,1}(\infty)$, the torque, 
$\Gamma_{m,n}$, is obtained.
Since we only consider the effect of the disk temperature gradient here,
we obtain $\Gamma_{m,n}^{(\beta)}$.

\begin{figure}[ht]
\includegraphics[width=8.4cm]{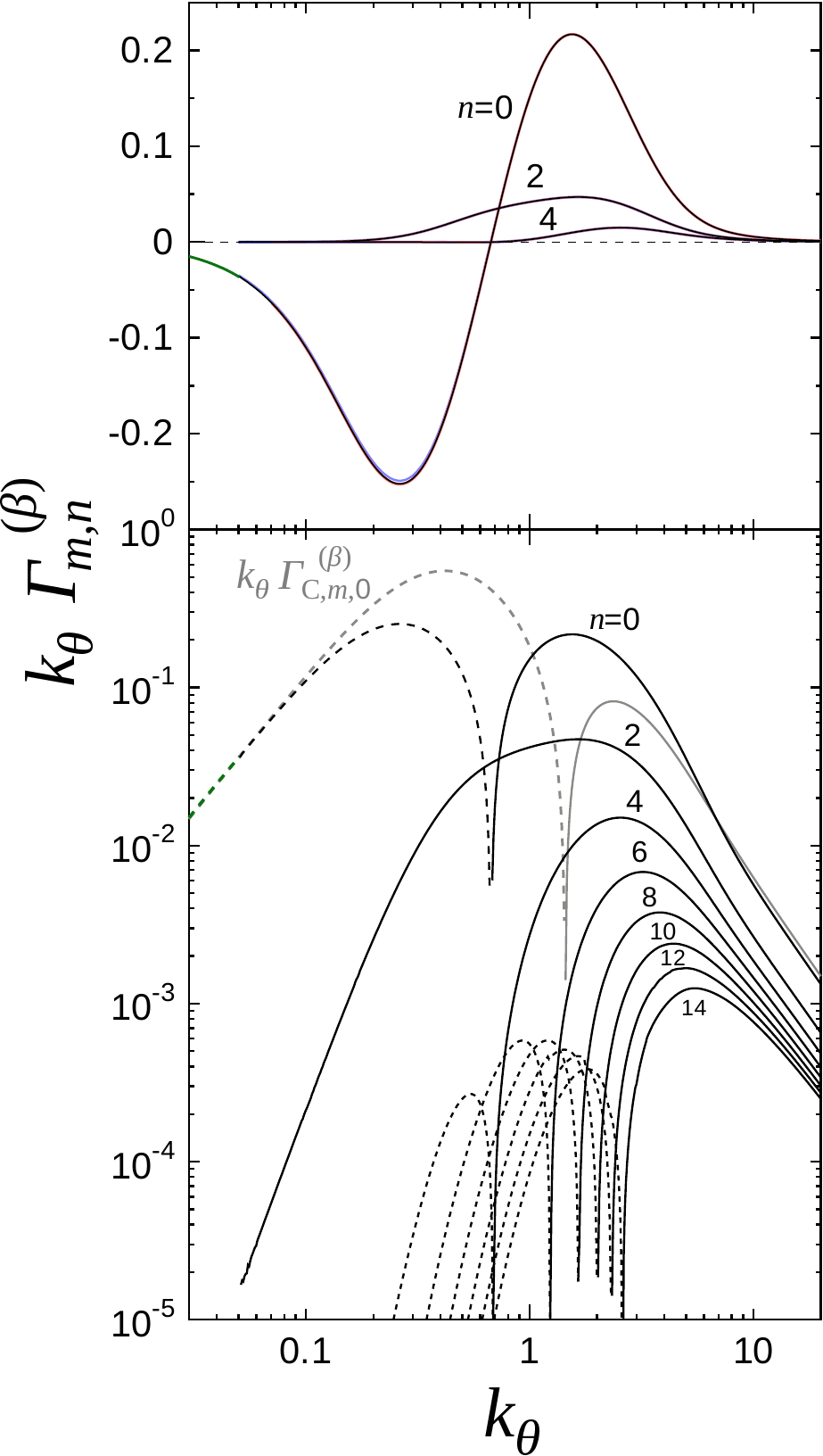}
\caption{Total torques, $\Gamma_{m,n}^{(\beta)}$, by each wave component $\eta_{m,n}$.
Here the torques due to the temperature gradient are plotted. The unit of the torques 
is $(h_p/r_p)\Gamma_0$. The top panel shows the total torques with $n=0, 2, 4$
for $\varepsilon = 10^{-3}$(blue), $10^{-4}$(black), and $10^{-5}$(red).
For $0.03 < k_\theta < 0.05$, the total torque with $n=0$ is obtained
with the method of Paper I and plotted by green lines.
The botom panel shows the total torques with $n=0$-$14$ for $\varepsilon = 10^{-4}$,
together with the corotation torques $\Gamma\sub{C,m,0}^{(\beta)}$ (gray).
The negative torques are plotted with dotted lines.
}
\label{fig:3}
%
\end{figure}
For the waves with different $m$ and $n$, similar integrations of the torque 
density give the torques, $\Gamma_{m,n}^{(\beta)}$, and the results are shown 
in Figure~\ref{fig:3}. 
In the top panel of Figure~\ref{fig:3}, we compare the torques with 
$\varepsilon = 10^{-3}, 10^{-4}$, and $10^{-5}$, for $n = 0$, 2, and 4. As 
expected from Figure~\ref{fig:2}, there is little dependence on $\varepsilon$. 
For $n$ = 0, the torque with $\varepsilon = 10^{-3}$ is slightly different from 
those with $\varepsilon = 10^{-4}$ and $10^{-5}$. Thus, the value of the torque 
is almost independent of $\varepsilon$ as long as $\varepsilon \le 10^{-3}$. 
For small $k_\theta$, it is difficult to solve the wave equation numerically
due to a small $\varepsilon$ of the Landau prescription.
For the two-dimensional waves with $n = 0$, fortunately, we can solve the wave 
equations without the Landau prescription as done in Paper I, because the wave 
equation with $n=0$ does not have the strongly divergent term. 
Thus, for $0.03 < k_\theta < 0.05$, the torque with $n=0$ is obtained by solving 
the wave equations with the method of Paper I instead of the Landau prescription, 
which is shown with a green line in Figure~\ref{fig:3}. The torques obtained by 
the two methods are almost continuous at $k_\theta$ = 0.05, confirming that both 
methods with and without $\varepsilon$ are consistent.

\begin{figure}
\includegraphics[width=8.4cm]{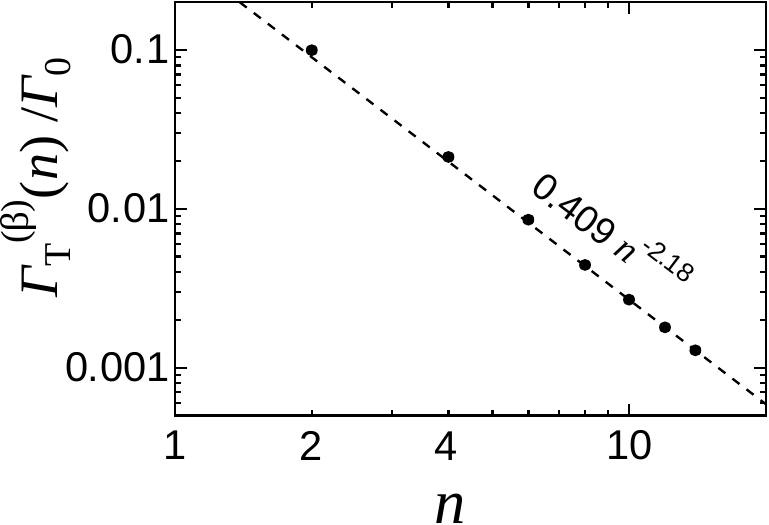}
\caption{The torques summed over $m$, $\Gamma^{(\beta)}\sub{T}(n)$, 
for $n \ge 2$. The dotted line represents the fitting formula for 
$n \ge 10$.}
\label{fig:4}
\end{figure}

The bottom panel of Figure~\ref{fig:3} shows the results with 
$\varepsilon = 10^{-4}$ for n=0 to 14 (solid lines are positive and dotted lines 
are negative). In addition, for $n=0$, the corotation torque given by 
Equation~(\ref{ctorque}) is also plotted. For $k_\theta < 0.1$, the corotation 
torque and the total torque are almost identical. This means that the Lindblad 
resonance is much weaker in this region. As $n$ increases, the peak 
value of each curve decreases but the decrease is not so rapid. As will be 
shown below, the contribution of larger $n$ is not negligible.

\begin{table*}
 \caption{Torques due to the radial temperature gradient, $\Gamma_T^{(\beta)}(n)$}
 \label{table:1}
 \centering
\begin{tabular}{cllllllllllll}
\hline \hline
$n$ 
& 0 & 2 & 4 &  6 & 8 & 10 &	12 & 14 
& sum($n\!\le\!14$) & sum($n\!>\!14$) & total\\
\hline
$\Gamma_T^{(\beta)}(n)$ 
& -0.1140 & 0.0997 & 0.0212 & 8.6e-3 & 4.4e-3 & 2.7e-3 & 1.8e-3 & 1.3e-3 
& 0.0258 & 7.0e-3 & 0.0328\\
\hline 
\end{tabular}
\tablecomments{The unit of the torques is $\Gamma_0$.}
\end{table*}
\begin{table*}
 \caption{Coefficients of the Corotation Torques}
 \label{table:2}
 \centering
\begin{tabular}{cccccc}
\hline \hline
n & $\Gamma\superscript{(curv)}\sub{CT}(n)$ 
& $\Gamma^{(\alpha)}\sub{CT}(n)$ &
$\Gamma^{(\beta)}\sub{CT}(n)$ & $\Gamma^{(\delta)}\sub{CT}(n)$ & 
$\Gamma^{(\mu)}\sub{CT}(n)$ \\
\hline
0 & -0.9457 & 0.6305 & -0.8575 & -- & -- \\
sum$(2\le n\le8)$ & -8.9e-3 & -0.5e-3 & -- & 2.7e-3 & 0.0274\\
\hline 
\end{tabular}
\tablecomments{The unit of the torques is $\Gamma_0$.}
\vspace{5mm}
\end{table*}
%
%
%
%

For $k_\theta \lesssim 0.05 $, we could not accurately obtain the wave solutions 
with $n \ge 2$. However, for the three-dimensional waves with such small $k_\theta$,
the torques are negligible compared to their peaks at $k_\theta \sim 1$. Therefore, 
we neglect the contribution of such small $k_\theta$ for the three-dimensional waves.

By integrating over $k_\theta$, we obtain $\Gamma_T^{(\beta)}(n)$ 
($=\sum_m \Gamma_{m,n}^{(\beta)}$) for $n = 0$ to 14. The integration range is 
$k_\theta =0.03$ to 20 for two-dimensional waves, while the contribution of small 
$k_\theta$ is neglected for three-dimensional waves. 
The obtained values are listed in Table~\ref{table:1} and also 
plotted in Figure~\ref{fig:4}. For $n > 4$, $\Gamma_T^{(\beta)}(n)$ is well 
fitted by a power-law function with the exponent of \mbox{-2.18}.
With the fitting formula,
we can extrapolate $\Gamma_T^{(\beta)}(n)$ for $n > 14$. The torque contribution of 
$n > 14$ estimated by the fitting formula is also shown in Table~\ref{table:1}. 
The contribution of $n > 14$ is $\simeq$ 20\%\footnote{
For other effects,  $\Gamma_T^{(X)}(n)$ also decrease with power-laws 
and similar extrapolations are done to obtain the summation of 
$\Gamma_T^{(X)}(n)$. These corrections are much smaller than $\Gamma_T^{(\beta)}(n)$.
}. 
Using this extrapolation, the total torque is finally obtained as
\begin{equation}
\sum_{n=0}^\infty \Gamma_T^{(\beta)}(n) = 0.0328 \Gamma_0.
\end{equation}
The summed corotation torque, $\Gamma\sub{CT}^{(\beta)}(0)$,
is obtained as \mbox{-0.8575}, which is much larger than
the total torque directly due to the temperature gradient, 
$\sum_n \Gamma_T^{(\beta)}(n)$.
This is because the Lindblad and corotation torques almost exactly 
cancel each other out for the effect of the temperature gradient. 
A similar cancellation between the Lindblad and corotation torques occur 
for the effect of the surface density gradient, which also results in 
the small total torques.

Note that the effect of the temperature gradient also appears indirectly 
through the effects of the pressure gradient and the scale height variation.
Since $h = \Omega/c_s$ and $p \propto T \Sigma /h$, the power-law indices
depend on $\beta$ as $\mu = 3/2 - \beta/2$ and 
$\delta = \alpha + 3/2 + \beta/2$, respectively. Thus, the torque term
proportional to $\beta$ is obtained as 
\begin{align}
\beta \frac{d\,  \Gamma\sub{total}}{d\beta}
&= \beta  
\left[ {\textstyle \sum\limits_n^\infty}\, \Gamma_T^{(\beta)}(n)
+ \frac12 {\textstyle \sum\limits_n^\infty}\, \Gamma_T^{(\delta)}(n)
- \frac12 {\textstyle \sum\limits_n^\infty}\, \Gamma_T^{(\mu)}(n)
\right] \nonumber \\
&= 0.4390 \, \beta \, \Gamma_0.
\label{torque_t}
\end{align}
The torque terms due to the other effects were obtained in Paper I. However, 
we have found some errors in the corotation torques of three-dimensional
waves. These errors in the corotation torques also affect the total torques.
We list the revised coefficients of the corotation torque 
in Table~\ref{table:2} 
as well as 
the torques due to the temperature gradient obtained in this paper. 
The revised coefficients of the total torque are given by
\begin{align}
&\sum\limits_{n=0}^{\infty} \Gamma\superscript{(curv)}\sub{T}(n) 
= 0.9439 \Gamma_0,
&\sum\limits_{n=0}^{\infty} \Gamma^{(\alpha)}\sub{T}(n) 
= -0.0337 \Gamma_0, \nonumber \\ 
&\sum\limits_{n=0}^{\infty} \Gamma^{(\delta)}\sub{T}(n) 
= 0.5703 \Gamma_0,
&\sum\limits_{n=0}^{\infty} \Gamma^{(\mu)}\sub{T}(n) 
= -0.2423 \Gamma_0.
\end{align}
These revised values are used in Equation~(\ref{torque_t}). Note that 
the torque contributions from the 3D waves in Table~\ref{table:2} are 
included in the Lindblad torque according to the new definition of the 
corotation and Lindblad torques although they belonged to the corotation 
torque in Paper I. 

Using these results of the linear calculations for each effect,
we obtain the formula of the total torque for locally isothermal disks as
\begin{align}
\Gamma\sub{total}/\Gamma_0 &= \, \, \, 1.436 + 0.537 \alpha + 0.439 \beta.
\label{total_torque}
\end{align}
The last term due to the temperature gradient is newly added in this study.
The terms due to the curvature and the surface density gradient
are slightly changed from the values of Paper I because of the errors in the
3D corotation torques in Paper I.
Note that this is the torque on the disk. The torque on the planet has the 
opposite sign. We can divide it into the Lindblad torque and the corotation 
torque. The corotation torque is given by the terms due to the two-dimensional 
waves in Table~\ref{table:2} and the Lindblad torque is given by
$\Gamma\sub{total}-\Gamma\sub{C}$. Thus we obtain
\begin{align}
\Gamma\sub{L}/\Gamma_0 &= \, \, \, 2.382 - 0.094 \alpha + 1.297 \beta, 
\label{Lindblad_torque}\\[1mm]
\Gamma\sub{C}/\Gamma_0 &=-0.630 \left( 3/2-\alpha \right) - 0.858 \beta.
\label{corotation_torque}
\end{align}
\vspace{1mm}
\section{Comparison with Previous Studies for Locally Isothermal Disks}
\subsection{Three-Dimensional Hydrodynamical Simulations}
\citet{2010ApJ...724..730D} performed a survey of three-dimensional
hydrodynamical simulations of type I planetary migration for locally
isothermal disks with different radial gradients of surface density and
temperature. They obtained the torque exerted by the planet on the disk as
\begin{align}
\Gamma\sub{total,DL10} &= \Gamma^{\alpha=\beta=0}\sub{total}
+ \frac{d\Gamma\sub{total}}{d\alpha} \alpha
+ \frac{d\Gamma\sub{total}}{d\beta} \beta \nonumber \\
&= (1.36 + 0.62 \alpha + 0.43 \beta)\Gamma_0.
\end{align}
Their value of $d\Gamma\sub{total}/d\beta$ is close to ours, $0.439\Gamma_0$. 
The difference is only 2\%. The other coefficients $\Gamma^{\alpha=\beta=0}\sub{total}$ 
and $d\Gamma\sub{total}/d\alpha$ have deviations of 5-15\% from our values. 
This difference {could} be due to the effect of the nonlinear horseshoe torque. 
We will discuss the horseshoe torque in Section 4.3.
\citet{2017MNRAS.471.4917J}
also examined the temperature gradient dependence of
the total torque with their 3D hydrodynamical simulations for locally isothermal disks. 
Their result is $d\Gamma\sub{total}/d\beta =(0.5\pm0.1)\Gamma_0$, which agrees with 
our value within their error bar. 
\citet{2010MNRAS.401.1950P}
suggested that the coefficient $d\Gamma\sub{total}/d\beta$ 
(or $d\Gamma\sub{C}/d\beta$) is not affected by the horseshoe torque in locally 
isothermal disks. The agreement in $d\Gamma\sub{total}/d\beta$ with 
hydrodynamical simulations supports Paardekooper~et~al.'s suggestion.

In the present study, we obtained the temperature-gradient term of the linear 
corotation torque as $\beta (d \Gamma\sub{C}/d\beta) = -0.858 \beta \Gamma$
in Equation~(\ref{corotation_torque}).
In \citet{2017MNRAS.471.4917J}, the corresponding term of the linear corotation 
torque is set to be $- \Gamma\superscript{lin}\sub{T} = - 1.0 \beta \Gamma_0$,
which should be replaced by our value, $-0.858 \beta \Gamma_0$.

\begin{figure}[t]
\includegraphics[width=8.5cm]{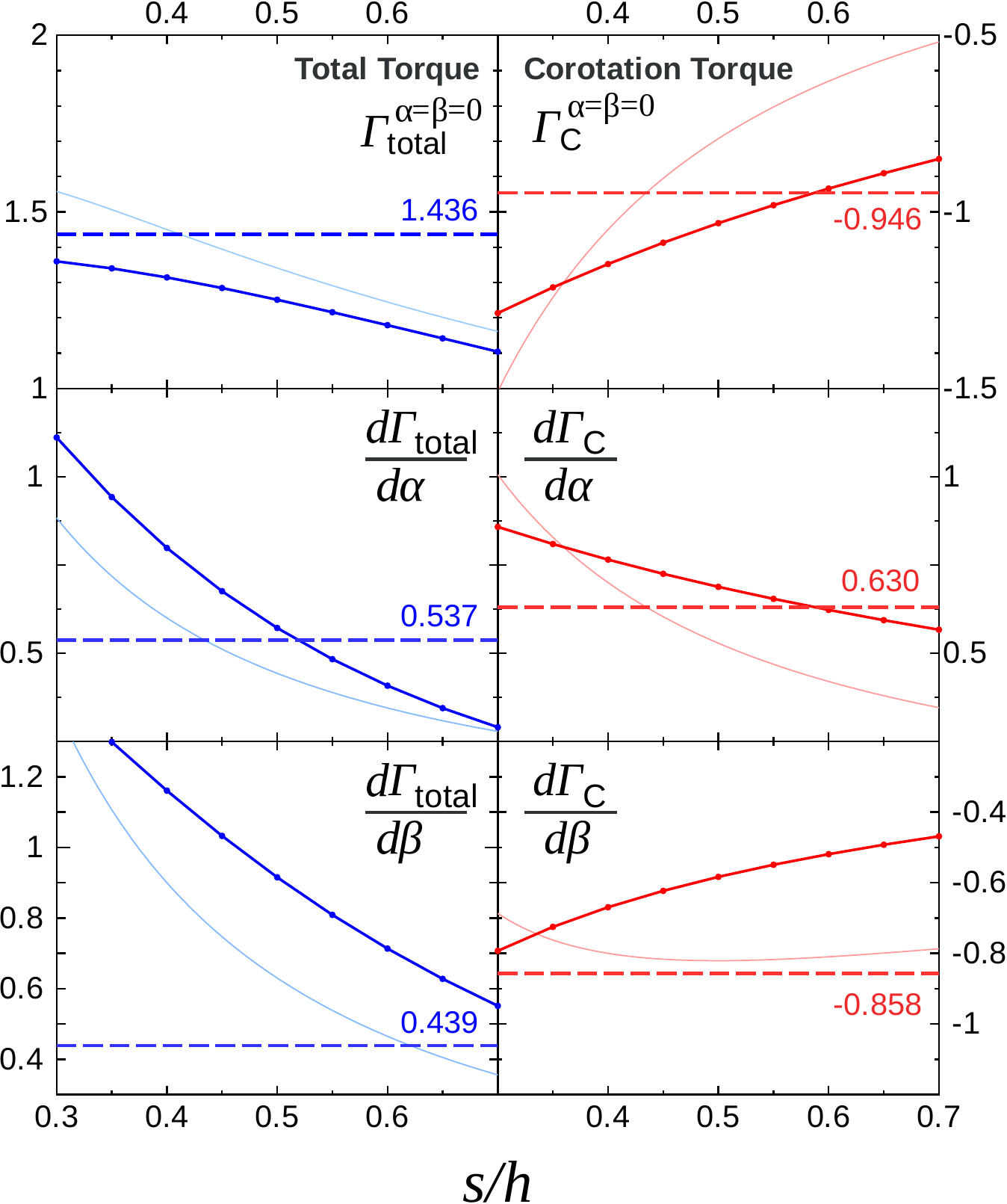}
\caption{Coefficients of the total torque and the corotation torque
in two- and three-dimensional linear calculations.
{The torque coefficients are plotted as functions of the smoothing length $s$.}
The light blue and light red lines show the coefficients in
the torque formulas of \citet{2010MNRAS.401.1950P}
for the locally isothermal case.
The horizontal broken lines represent the 3D linear results
of the present paper. Our 2D linear results are plotted with
the blue and red lines.}
\label{fig:5}
%
\end{figure}

\subsection{Two-Dimensional Linear Calculation by Paardekooper et al. (2010)}
\citet{2010MNRAS.401.1950P}
constructed the torque formulas based on their
2D linear calculations. Figure~\ref{fig:5} shows a comparison of our 3D 
total torque and corotation torque with Paardekooper et al.'s torque formulas 
for locally isothermal disks. Since their 2D torque formulas depend on the
smoothing length included in the gravitational potential, Figure~\ref{fig:5} 
shows each torque coefficient as a function of the smoothing length. 
Furthermore, we also performed 2D linear calculations and plotted the results 
as well. Although the torque formulas of \citet{2010MNRAS.401.1950P}
are expressed by simple power-law fits, their formulas are almost in agreement 
with our 2D linear calculations. The torque coefficients of the 3D linear 
calculations agree with Paardekooper et al's formulas with a smoothing length 
of $\simeq 0.4h$ for both the curvature term and the surface density gradient
term of the total and corotation torques. 
On the other hand, the temperature gradient term obtained in the present 
3D calculation agrees with the formula with $s \simeq 0.6h$ for the total 
torque, while a good agreement between the 3D result and their formula is 
seen in the wide range of s=0.4-0.7h for the corotation torque. 
These comparisons indicate that it is difficult to
reproduce all the dependencies in the 3D torques with the 2D torque formulas
with a single smoothing length.

\subsection{Horseshoe Torque in Three-Dimensional Disks}
In hydrodynamical simulations, the torque contribution around corotation
is often closer to the horseshoe torque described below than to the 
linear corotation torque because of nonlinear effects,
though it depends on the saturation \citep[e.g.,][]{2010ApJ...723.1393M,2011MNRAS.410..293P}.
For 2D isothermal disks, the unsaturated horseshoe torque on the planet, 
$\Gamma\sub{C,hs}$, is given by \citep[e.g.,][]{1991LPI....22.1463W, 2001ApJ...558..453M}

\begin{equation}
\Gamma\sub{C,hs} =
\frac34 \left( \frac32 - \alpha \right) \sigma_p \Omega_p^2 x\sub{hs}^4,
\label{hs_torque}
\end{equation}
where $x\sub{hs}$ is the half-width of the ring-like horseshoe region.
\citet{2006ApJ...652..730M} showed that the half-width $x\sub{hs}$
is given by
\begin{equation}
x\sub{hs} =
\left(-\frac83 \eta'\sub{S}\right)^{1/2}\Omega_p^{-1}
\end{equation}
for 2D disks. In this equation, the suffix S indicates that the perturbation,
$\eta' (=\eta + \phi_p)$, is evaluated at the stagnation point of the
horseshoe flow. Since $\eta'\sub{S} \sim \mbox{O}(\lambda r_p/h_p)$, the 
horseshoe torque is obtained as $A (3/2-\alpha)\Gamma_0$, where $A$ is a
coefficient of the order of unity. The half-width of the horseshoe region,
$x\sub{hs}$, can also be obtained from the streamlines of the horseshoe
flow in hydrodynamical simulations. In 2D disks, $x\sub{hs}$ and the horseshoe
torque depend on the smoothing length 
\citep[e.g.,][]{2006ApJ...652..730M,2010ApJ...723.1393M,
2009MNRAS.394.2297P, 2010MNRAS.401.1950P}.
Note that in completely isothermal disks, the horseshoe torque 
is proportional to the vortensity gradient (i.e., the factor, $3/2-\alpha$)
as well as the linear corotation torque of Equation~(\ref{corotation_torque}).

\citet{2016ApJ...817...19M} performed 3D isothermal simulations to
obtain the horseshoe torque. Their 3D simulations show that the horseshoe 
flow is almost two-dimensional and nearly independent of $z$. 
Thus, Equation~(\ref{hs_torque}) for the horseshoe torque in 
the two-dimensional case can apply to the three-dimensional case. 
From their 3D simulations, they obtained the width of the horseshoe region as
\begin{equation}
x\sub{hs} = 1.05
\left(q\frac{r_p}{h_p}\right)^{\!1/2} \!r_p,
\end{equation}
where $q = M_p/M_*$, and the horseshoe torque is 
\begin{equation}
\Gamma\sub{C,hs} = 0.91 \left( \frac32 - \alpha \right)
\Gamma_0.
\label{horseshoe_masset}
\end{equation}
%
This value of the horseshoe torque for 3D isothermal disks
is consistent with the torque formula by 
\citet{2010MNRAS.401.1950P} if the smoothing length of $\simeq 0.5h$ is used.
Moreover, it is about 1.4 times as large as the vortensity term of our linear 
corotation torque of Equation~(\ref{corotation_torque}).
Note that the simple horseshoe torque formula of 
Equation~(\ref{horseshoe_masset}) is valid only for low-mass planets 
with $q \le 0.1 (h_p/r_p)^3$ \citep{2017MNRAS.471.4917J}.

Here we also obtain the horseshoe torque by estimating $\eta'\sub{S}$ with the 
3D linear calculation. Since the horseshoe flow is shown to be two-dimensional, $\eta'$ is 
estimated using the two-dimensional waves with $n=0$, and the three-dimensional waves 
with $n > 0$ are negligible. Thus, $\eta'$ is obtained by summing only the terms for 
$n=0$ over $m$ in Equation~(\ref{eta2}) for the horseshoe region. Since the stagnation point 
is approximately at the corotation radius, we take the minimum value of $\eta'$ at 
$x = 0$ as $\eta'\sub{S}$. In the summation, the term for $m = 0$ is omitted. In 
addition, the term for $m = 1$ originally contains an indirect term, which is also 
ignored in the modified local approximation. 

For comparison with \citet{2016ApJ...817...19M},
we set $h_p/r_p = 0.05$. We also include the first-order terms of the modified local
approximation. In the first-order terms, we include the curvature terms and set 
$\alpha = 3/2$, $\beta = 0$, and $\delta = 3/2 + \alpha = 3$ as in the nominal case 
of \citet{2016ApJ...817...19M}.
As a result, we obtain
\begin{align}
&\eta'\sub{S} =-0.399 \, q \frac{r_p}{h_p} r_p^2 \Omega_p^2, \qquad
x\sub{hs} = 1.032 \sqrt{q\frac{r_p}{h_p}} r_p, 
\nonumber\\
&\Gamma\sub{C,hs} = 0.849 \left( \frac32 - \alpha \right)
\Gamma_0.
\label{hs-lin}
\end{align}
This is close to the result of \citet{2016ApJ...817...19M},
with a deviation of several \% in the horseshoe torque. Thus, the linear 
calculation gives a good approximation of the horseshoe torque. 

The results of Equation~(\ref{hs-lin}) are 
derived using the parameter set of \citet{2016ApJ...817...19M}.
Note that the 
numerical coefficients in Equation~(\ref{hs-lin}) depend on the parameters. To 
obtain $\eta'\sub{S}$ above, we included the first-order terms of the modified local 
approximation, which depend on the values of $\alpha$ and $\delta$. If we ignore the 
first-order terms and use the shearing sheet model, the horseshoe torque is 3\% 
smaller than Equation~(\ref{hs-lin}). The normalized horseshoe torque 
$\Gamma\sub{C,hs}/\Gamma_0$ also depends on the disk aspect ratio $h_p/r_p$; for 
$h_p/r_p = 0.03$, $\Gamma\sub{C,hs}/\Gamma_0 = 0.952(3/2 - \alpha)$, which is 
about 10\% larger than Equation~(\ref{hs-lin}).
The similar dependence of the horseshoe torque (or the width) on $h_p/r_p$ is
seen in the simulation results of \citet{2017MNRAS.471.4917J} (in their Fig.~6).

In locally isothermal disks, the horseshoe torque also has the 
term related to the radial temperature gradient.
In \citet{2017MNRAS.471.4917J}, such a term of the unsaturated horseshoe torque 
on the planet is given by 
\begin{equation}
\Gamma\superscript{UHD}\sub{T} = 0.73 \beta \Sigma_p \Omega_p x\sub{hs}^4 
= 0.89 \beta \Gamma_0
\end{equation}
for a low-mass planet.
This is also close to our temperature-gradient term of the linear corotation torque
of Equation~(\ref{corotation_torque}). This agreement between the horseshoe and 
linear corotation torques in locally isothermal disks has already been suggested 
by Paardekooper et al. (2010).

\section{Sumamry}

In the present study, we have performed linear calculations to determine the torques 
of Type I planetary migration for 3D locally isothermal disks with radial temperature 
gradients, whereas Paper I gave the torques for completely isothermal disks. In a disk 
with a radial temperature gradient, the wave equation has a term that strongly diverges 
at the corotation, which {\bf causes a non-removal singularity in the wave solution.}
As a result, 
obtaining the wave solutions was challenging. We obtained the wave solutions, 
suppressing the divergence with the Landau prescription. Our results are summarized as 
follows.

\begin{itemize}

\item[1.] 
The waves excited by a planet have a large amplitude near the corotation due to 
the divergent term, and the degree of amplification depends on the parameter of 
the Landau prescription (Fig.~\ref{fig:1}). Our wave solutions show that this wave 
amplification at corotation does not affect the region away from corotation. This 
limited influence of the divergent term is explained by the fact that the waves 
propagate towards corotation and dissipate at corotation. We also found that this 
wave amplification does not affect the torque on the planet due to the very narrow 
range of the wave amplification (Fig.~\ref{fig:2}). Thus, we can obtain a definite 
torque even though the wave equation has {a divergent term of the third pole}
(Fig.~\ref{fig:3}).

\item[2.]
The total torque was obtained by summing the torque contributions of each wave 
component. The obtained torque directly due to the temperature gradient is smaller 
than the torque due to other effects obtained in Paper I. Thus, the torque due to the
temperature gradient mainly results from the pressure gradient and the scale height 
variation indirectly. The total torque considering all effects, including the 
temperature gradient, is given by Equation~(\ref{total_torque}).

\item[3.]
Equation~(\ref{ctorque}) shows the revised formula of the corotation torque for 
locally isothermal disks with temperature gradients. Due to the new definition of
corotation torque, three-dimensional waves do not contribute to the corotation torque. 
The revised formula gives the corotation torque as Equation~(\ref{corotation_torque}), 
and the Lindblad torque is given by Equation~(\ref{Lindblad_torque}).

\item[4.]
The torque due to disk temperature gradient obtained by our 3D linear calculation 
agrees well with the result of the previous 3D hydrodynamical simulations for 
locally isothermal disks \citep{2010ApJ...724..730D,2017MNRAS.471.4917J}. 
The torque formula of \citet{2010MNRAS.401.1950P} based on 2D linear calculations 
requires a large smoothing length of 0.6h to make the torque due to the 
temperature gradient consistent with the 3D result. Therefore, it is difficult to 
reproduce all the dependence of the 3D torque on each effect using a 2D calculation 
with a single smoothing length.

\item[5.]
Equation~(\ref{hs-lin}) shows the horseshoe torque obtained through our 3D linear 
calculation. This result is also close to that of 3D hydrodynamical simulations
by \citet{2016ApJ...817...19M}.
The temperature gradient term of the horseshoe torque is very close
to that of our linear corotation torque for locally isothermal disks.

\end{itemize}
\vspace*{1mm}

\noindent
The authors thank the anonymous referee for his/her insightful suggestions.
The authors also thank Fr\'{e}d\'{e}ric Masset and Takayuki Muto for thier valuable comments.
One of the authors (HT) would like to express gratitude to Bill Ward for his 
patient encouragement during the long and unsuccessful attempt to solve the divergence 
problem at corotation. 
This study was supported by JSPS KAKENHI grant Nos. 22H01278, 19K03941, and 18H05438.

%




\appendix

\section{Excitation and Propagation of Density Waves}
The excitation and propagation of density waves in two- and 
three-dimensional disks can be explained in terms of dispersion relations, 
and of course, such an explanation has already been given by previous 
studies \citep[e.g.,][]{linshu68, 1979ApJ...233..857G,1993ApJ...409..360L,
1998PASJ...50..141T}. Here we briefly review their results on 
non-self-gravitating gaseous disks. In the WKB approximation, the radial 
dependence of the Fourier-Hermite components is written in the form 
$\exp(i\int^r k_r dr)$ and substituted into 
Equations~(\ref{eulereq_r})-(\ref{eulereq_z}) to obtain the following 
dispersion relation (\citealt{1998PASJ...50..141T})\footnote{
\citet{1993ApJ...409..360L} obtained a similar dispersion relation 
for 3D axisymetric waves though they also include the buoyancy using 
the polytropic equation of state.}.
\begin{equation}
k_r^2 = \frac{D_1 D_n}{c^2 m^2(\Omega_K -\Omega_p)^2},
\label{dispersion}
\end{equation}
where $D_n =  n \Omega_K^2 - m^2(\Omega_K -\Omega_p)^2$. In the above 
derivation, we have neglected the terms proportional to $1/r$ in the 
WKB approximation and also omitted the terms proportional to the small 
$\Delta \Omega$. As a result, Equation~(\ref{dispersion}) agrees with 
the dispersion relation in the shearing sheet model. The approximated 
dispersion relation is valid for studying the qualitative properties of 
wave excitation and propagation. 
\begin{figure}[b]
\includegraphics[width=8cm]{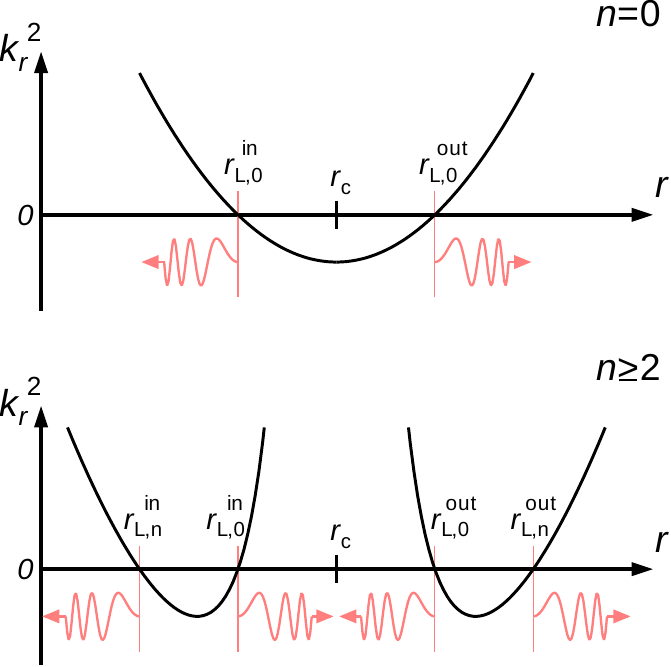}
\caption{The $r$-dependence of the square of the wavenumber $k_r^2$
for two-dimensional waves with $n=0$ (upper panel) and three-dimensional
waves with $n\ge 2$ (lower panel).
Density waves are excited at the Lindblad resonances, where $k_r=0$,
and propagate into the regions where $k_r^2>0$.
}
\label{figA}
\end{figure}

The $r$-dependence of $k_r^2$ is illustrated in Figure~\ref{figA}.
In the case of two-dimensional density waves ($n=0$), the wavenumber 
$k_r$ vanishes at the (original) inner and outer Lindblad resonances,
$r_{L,0}^{\, \mbox{\footnotesize in}}$, $r_{L,0}^{\, \mbox{\footnotesize out}}$,
where $D_1=0$ or $\Omega_K = \frac{m}{m\pm 1} \Omega_p$.
Density waves are excited at these Lindblad resonances and propagate
away from the planet. The propagation regions have a positive $k_r^2$.
As the waves propagate far away from the resonances, both the
wavenumbers $|k_r|$ and the amplitudes increase. As a result, the density 
wave decays with shock dissipation \citep{2001ApJ...552..793G}.
In the region between the two Lindblad resonances, $k_r^2$ is negative and thus 
the waves are evanescent. The corotation point, $r_c$, and the nearby radial 
position of the planet, $r_p$, are almost in the center of this region.

For three-dimensional waves with $n \ge 2$, $k_r^2$ is positive even 
near corotation. The wavenumber also vanishes at the new inner and 
outer Lindblad resonances, $r_{L,n}^{\, \mbox{\footnotesize in}}$, 
$r_{L,n}^{\, \mbox{\footnotesize out}}$, where $D_n=0$ (or 
$\Omega_K(r) = \frac{m}{m\pm \sqrt{n}}\Omega_p$).
The new Lindblad resonances are farther away from the planet (or $r_c$) 
than the original Lindblad resonances. The three-dimensional waves with 
each $n$ mode are excited at these four Lindblad resonances. The waves 
excited at the far-side Lindblad resonances propagate {in the far region
where $D_n <0$,}
while the waves excited at the near-side Lindblad resonances 
propagate towards corotation {in the near region where $D_1 >0$.}
{The former waves correspond to the p-mode with high frequencies as well as
the two-dimensional waves due to the original Lindblad resonance, and
the latter waves correspond to the g-mode (or the r-mode) with low frequencies 
\citep{1993ApJ...409..360L, 1998ApJ...504..983L}.}
Since the wavenumber is divergent at corotation, 
the latter waves are dissipated in the neighborhood of corotation. 

The dispersion relation (\ref{dispersion}) also gives the radial 
group velocity of the waves, $v_{g,r}$. Since the angular frequency 
of the waves is given by $m\Omega_p$ in Euqation~(\ref{eta2}), 
we obtain
\begin{equation}
v_{g,r} = m \frac{d\Omega_p}{d k_r} 
= \frac{c^2 k_r}{m(\Omega_K-\Omega_p)}
\left[
\frac{n\Omega_K^4}{m^4(\Omega_K-\Omega_p)^4}-1
\right]^{-1}.
\end{equation}
The sign of $v_{g,r}$ determines the direction of radial propagation.
We note that the directions of propagation in Figure~\ref{figA}
are consistent with the signs of $v_{g,r}$ in each wave region when 
the wavenumber $k_r$ is positive (i.e., the waves are trailing).
In fact, trailing density waves are obtained in our linear 
calculations (see Figure~\ref{fig:1} as well as in the previous 
three-dimensional linear calculations (\citealt{1998PASJ...50..141T};
Paper I). This is obtained by applying the radiation boundary condition.


\bibliography{ms}{}
\bibliographystyle{aasjournal}



\end{document}